\documentclass[10pt,twocolumn,twoside]{IEEEtran}

\ifCLASSINFOpdf
\else
\fi
\usepackage{lipsum}% http://ctan.org/pkg/lipsum

\usepackage{subfigure}
\usepackage{amsmath,epsfig}
\usepackage{amsxtra, color}
\usepackage{amsmath}
\usepackage{amsthm}
\usepackage{amsfonts}
\usepackage{algorithm}
\usepackage{graphicx}
\usepackage{algorithmic}
\usepackage{amsthm}
\usepackage{hyperref}
\usepackage[noadjust]{cite}
\usepackage{color}
\usepackage{caption}
\usepackage{amssymb}
\usepackage{mathtools}
\usepackage{pstricks}
\usepackage{pst-node}
\usepackage{pst-grad}
\usepackage{pst-plot}
\usepackage{pstricks-add}

\begin{document}
\title{Transmit Beamforming for Interference Exploitation in the Underlay Cognitive Radio Z-channel}

\author{Ka Lung Law,
\thanks{
Ka Lung Law, and Christos Masouros are with Department of Electronic and Electrical Engineering - University College London, Torrington Place, London WC1E 7JE, UK  (E-mail: k.law@ucl.ac.uk; chris.masouros@ieee.org).%; kai-kit.wong@ucl.ac.uk).

Marius Pesavento is with the Communication Systems Group, Technische Universit\"at Darmstadt, D-64283 Darmstadt, Germany (e-mail: mpesa@nt.tu-darmstadt.de).

The work of C. Masouros was supported by the Royal Academy of
Engineering, UK and the Engineering and Physical Sciences Research Council
(EPSRC) project EP/M014150/1.
%Gan Zheng is with School of Computer Science and Electronic Engineering, University of Essex, Colchester, CO4 3SQ, UK (E-mail: ganzheng@essex.ac.uk). He is also affiliated with Interdisciplinary Centre for Security, Reliability and Trust (SnT), University of Luxembourg, Luxembourg.
}
Christos Masouros,
Marius Pesavento, 
%Kai-Kit Wong, {\it Senior Member, IEEE}
%Gan Zheng, {\it Senior Member, IEEE}

%\IEEEauthorblockN{Ka Lung Law, Xin Wen, Minh Thanh Vu, and Marius Pesavento, Member, IEEE} %\\
%\IEEEauthorblockA{Darmstadt University of Technology, Merckstr. 25, D-64283 Darmstadt, Germany}
}

\maketitle

\begin{abstract}
	This paper introduces novel transmit beamforming approaches for the cognitive radio (CR) Z-channel. The proposed transmission schemes exploit non-causal information about the interference at the SBS to re-design the CR beamforming optimization problem. This is done with the objective to improve the quality of service (QoS) of secondary users by taking advantage of constructive interference in the secondary link. The beamformers are designed to minimize the worst secondary user's symbol error probability (SEP) under constraints on the instantaneous total transmit power, and the power of the instantaneous interference in the primary link. The problem is formulated as a bivariate probabilistic constrained programming (BPCP) problem. We show that the BPCP problem can be transformed for practical SEPs into a convex optimization problem that can be solved, e.g., by the barrier method. A computationally efficient tight approximate approach is also developed to compute the near-optimal solutions. Simulation results and analysis show that the average computational complexity per downlink frame of the proposed approximate problem is comparable to that of the conventional CR downlink beamforming problem. In addition, both the proposed methods offer significant performance improvements as compared to the conventional CR downlink beamforming, while guaranteeing the QoS of primary users on an instantaneous basis, in contrast to the average QoS guarantees of conventional beamformers. 
	
\end{abstract}
\IEEEpeerreviewmaketitle

\section{Introduction}

Dynamic spectrum access (DSA) in cognitive radio (CR) networks 
has provided an effective way to increase the radio resource utilization and spectral efficiency, by allowing the utilization of the licensed spectrum by secondary links \cite{Mitola1999,Haykin2005,Qing2007,Sankaranarayanan2005,Zhang2010}.
%With a existing primary network, the usage of secondary spectrum has been provided an efficient way to solve the radio resource scarcity. 
In underlay CR networks, the primary users (PUs) have the highest priority to access the spectrum without being aware of the existence of the unlicensed secondary user (SU) network. However, under the underlay CR paradigm the PU network is willing to grant spectrum access to the SU network under the premise that 
the interference created by the secondary base station (SBS) does not exceed a predefined threshold \cite{Sankaranarayanan2005}. 
With the knowledge of channel state information
(CSI) for both the PUs and SUs at the SBS, a fundamental challenge for CR is to enable
opportunistic spectrum access while meeting the quality of service (QoS) requirements of the SUs, e.g., in terms of signal
to interference plus noise ratio (SINR), system capacity, or
symbol error rate (SER). The policy of CR network in return
guarantees to protect the PUs from interference induced by
the SUs \cite{Zhang2010,Zhang2008,Islam2007}.%The coexistence of PUs and SUs in CR network through spectrum sharing can improve the overall spectral by allowing SU network transmissions \cite{Zhang2010,Zhang2008,Islam2007}.} %With the knowledge of channel state information (CSI) for both the PUs and SUs at SBS, a fundamental challenge for CR is to enable opportunistic spectrum access by exploiting spectrum to meet the QoS requirements of the SUs, e.g., in terms of signal to interference plus noise ratio (SINR), system capacity, or symbol error rate (SER). The policy of CR network in return guarantees to protect the PUs from interference induced by the SUs \cite{Zhang2010,Zhang2008,Islam2007}. 

To facilitate the utilization of the available radio spectrum, CR employs techniques from traditional (non-CR) wireless networks \cite{Zhang2010}. Existing studies in the traditional networks have shown that the QoS can be improved by exploiting the spatial domain with the use of multiple antennas at the SBS \cite{dahlman20114g,wirelessbook}. Several beamforming techniques haven been developed for the conventional wireless downlink to amplify the signal and suppress the interference by exploiting the CSI \cite{dahlman20114g,wirelessbook,Rashid1998,schubert2004solution,bengtsson1999optimal, bengtsson2001optimal, Shamai2006, gershman2010convex, Choi2015}. With the introduction of pre-coding techniques, multiuser downlink designs have been developed extensively in non-CR wireless communications.  
Dirty paper coding (DPC) techniques have been introduced to pre-eliminating potential interference experienced at the receiver already before transmission \cite{Costa1983, Erez2005}. However, the DPC techniques, despite being capacity optimal, involve non-linear and non-continuous optimization, which require sophisticated search algorithms and assume the data are encoded by codewords with infinite length \cite{Hassibi2002}. Several heuristic approaches are proposed to reduce the complexity \cite{Windpassinger2004,Masouros2012a,Rodriguez2014}. Nevertheless, they are generally far from being practical in current communication standards due to high computational complexity. %As the PUs have priority in the spectrum access, the beamforming designs in the CR network generally have to fulfill the additional requirement that the average interference power created by SUs at the primary users must reside below a predefined threshold. In recent years, several beamforming techniques have been proposed that take the interference power created at the PUs into account. 

As regards the CR transmission, the power minimization and SINR balancing problem for SUs with average interference power constraints of the primary users has been discussed in \cite{Fu2009,Zhang2010}. 
Conventionally this problem is solved by (sequential) approximation as of second-order cone programs (SOCPs).  %In \cite{Kim2011}, Kim et al propose a decentralized approach to maximize the user weighted sum-rate of the CR network. When there is a single SU, the ergodic capacity subject to a power constraint was derived \cite{Ghasemi2007}. A game-theoretic approach for CR networks has been proposed in \cite{Niyato2008,Maskery2009}. 
%The CR beamforming techniques that provide worst-case robustness against the channel uncertainty have been studied in \cite{Wajid2013,Singh2014}. The worst-case robust beamforming problem is generally solved suboptimally using the semidefinite relaxation (SDR) technique along with randomization. Several advance techniques go beyond SDR with randomization, such as general rank beamforming \cite{Law2015a}, and stochastic transmit beamforming \cite{Wu2013}. 
To achieve more flexibility than that of the worst-case based design, channel outage univariate probabilistic constrained programming (UPSP) downlink beamforming problem has been developed \cite{Wajid2013,Chalise2007}. Nevertheless, the techniques of solving for UPSP problem could not be extended to multivariate probabilistic constrained programming problem as the problem is non-convex in general \cite{Prekopa1970}. 

In order to improve the performance, the above mentioned SINR-based CR downlink beamforming problems are designed to mitigate the multiuser interference among the SUs. However, the associated drawback is that in SINR-based designs, some degrees of freedom in the beamforming design are used to suppress and eliminate the interference, which results in an overall increase of the transmitted power. Moreover, with conventional CR beamforming \cite{Fu2009,Zhang2010}, which only constrains the average interference, the instantaneous interference at the PUs at individual time instants may largely exceed the predefined thresholds. This can be overcome by utilizing the knowledge of both CSI and SU's information symbols at the SBS to exploit the resulting interference in the secondary links. In this case beamformers can be designed to enhance the useful signal by steering the received signals, containing both the desired and the interfering signals, into the correct detection region instead of separately amplifying and suppressing the desired and the interfering signals, respectively \cite{Masouros2010a,Masouros2010,Masouros2011,Masouros2012a,Masouros2012b,Masouros2013,Zheng2014,Masouros2015,Alodeh2015}. This approach is also known as a constructive interference precoding. Closed-form linear and non-linear constructive interference precoders were developed for the non-CR downlink to achieve higher SINRs at the receivers without the requirement of additional transmit power as compared to the interference suppression techniques \cite{Masouros2010a,Masouros2010,Masouros2011,Masouros2012a,Masouros2012b,Masouros2013,Zheng2014}.  
To further reduce the transmit power, the beamforming optimization-based precoders are discussed in \cite{Masouros2015,Alodeh2015,Law2015}. %One form of the optimization problem is to minimize the transmitted power by constructing useful interference to enhance the signal and at the same instant guaranteeing the QoS for all users.

The above works, build upon the observation that in time division duplexing (TDD) systems, downlink channels CSI can be obtained from uplink training due to the assumption of uplink-downlink channel reciprocity \cite{Smith2004}. Thus the training process can be simplified without the feedback of CSI estimate from the receiver. More importantly, by using the constructive interference technique the decoding process can be further simplified. That is, with conventional beamforming the receiver needs to calculate the composite channel (composed by the product of its downlink channel with the corresponding beamformer) for equalization and detection. With the proposed technique however, it will be shown that, as the received symbols fall in the constructive region of the signal constellation, no such equalization is required at the receiver and a simple decision stage at the receiver side suffices. Accordingly, The benefit for the proposed scheme is threefold: 1.) There is no need to send common pilots to the users to estimate the MISO channels. 2.) There is no need to signal the beamformers for the users compute the composite channels for equalization. 3.) It is not subject to the associated errors in the composite channel due to the estimation errors and CSI quantization during the feedback procedure, which further deteriorate the performance.

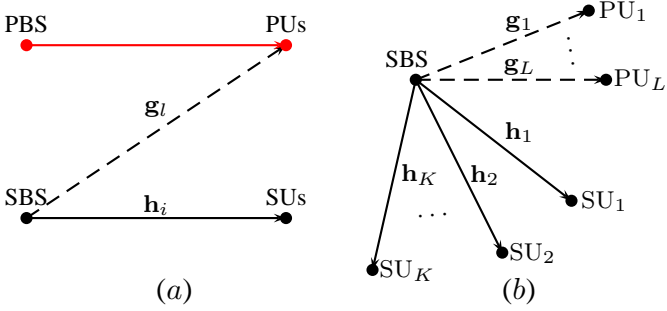
\begin{figure}[t]
	\centering
	\psscalebox{1.15}{
		\begin{pspicture}(7,4)
		
		\psset{blendmode=2}
		
		\psset{linewidth=1.0pt}
		
		\def\radius{1.0}       % Radius of L1-Ball
		
		\def\Theta{60 } \def\Radius{4 }
		
		% Axis
		%	\psset{arrowsize=3.5pt,arrowinset=0.02}
		%	\psline[linewidth=0.7pt]{->}(0,1.5)(3,1.5)
		%	\psline[linewidth=0.7pt]{->}(0,0.0)(0,4.5)
		
		%	\psline[linecolor=black,linewidth=0.7pt,linestyle=dashed]{-}(0.0,1.5)(3,4.5)
		
		%	\pspolygon*[linecolor=black!80!white,opacity=0.3](0.4,1.9)(1.4,4.9)(3.4,2.9)
		%	\pspolygon*[linecolor=black!80!white,opacity=0.3](0,1.5)(1,4.5)(3,2.5)
		
		\psline[linecolor=black,linewidth=0.7pt]{->}(0,1)(3,1)\rput(1.5,1.15){\footnotesize${\mathbf h}_i$}
		\psline[linecolor=black,linestyle=dashed,linewidth=0.7pt]{->}(0,1)(3,3) \uput[90]{0}(1.5,2){\footnotesize${\mathbf g}_l$}
		\psline[linecolor=red,linewidth=0.7pt]{->}(0,3)(3,3)
		\pscircle*(0,1){0.07}
		{\rput(0,1.25){\footnotesize SBS}}
		\pscircle*(3,1){0.07}
		{\rput(3,1.25){\footnotesize SUs}}
		\pscircle*[linecolor=red](0,3){0.07} 
		{\rput(0,3.25){\footnotesize PBS}}
		\pscircle*[linecolor=red](3,3){0.07} 
		{\rput(3,3.25){\footnotesize PUs}}

		%	\pscircle*(0.707,2.207){0.07} %\uput[90]{45}(0.707,3.4){$b_i$}		

		%	\psline[linecolor=black,linewidth=0.7pt]{->}(2.4,2.9)(2.8,3.3)

		% Axis
		%	\psset{arrowsize=3.5pt,arrowinset=0.02}
		%	\psline[linewidth=0.7pt]{->}(4,1.5)(7,1.5)
		%	\psline[linewidth=0.7pt]{->}(4,0.0)(4,4.5)
		
		%	\pspolygon*[linecolor=black!80!white,opacity=0.3](4,1.5)(5,4.5)(7,2.5)
		%	\pscircle*(4.707,2.207){0.07} \uput[0]{45}(4.6,2.1){\footnotesize$b_i$}		
		
		%	\psline[linecolor=black,linewidth=0.7pt,linestyle=dashed]{-}(4.0,1.5)(7,4.5)

		%	{\color{red}\uput[90]{45}(5,3.4){\footnotesize$y_i$} \uput[90]{45}(4.9,2.25){\footnotesize$\psi_i$}}
		%	\psline[linecolor=red,opacity=0.9,linewidth=1pt]{->}(4.0,1.5)(5,3.5)
		%	\psarc[linecolor=red,linewidth=1.0pt](4.0,1.5){1.2}{45}{63.435}
		
		%	\psarc[linewidth=1.0pt](4.0,1.5){0.8}{45}{71.565}
		%	\uput[90]{45}(4.35,1.7){\footnotesize$\theta$}

		\psline[linecolor=black,linestyle=dashed,linewidth=0.7pt]{->}(4.5,2.6)(6.5,3.4) \rput(5.7,3.25){\footnotesize${\mathbf g}_1$}
		\psline[linecolor=black,linestyle=dashed,linewidth=0.7pt]{->}(4.5,2.6)(6.7,2.6)\rput(5.7,2.75){\footnotesize${\mathbf g}_L$}
		\psline[linecolor=black,linewidth=0.7pt]{->}(4.5,2.6)(6.3,1.2) \rput(5.7,2.0){\footnotesize${\mathbf h}_1$}
		\psline[linecolor=black,linewidth=0.7pt]{->}(4.5,2.6)(5.5,0.6)	\rput(5.3,1.5){\footnotesize${\mathbf h}_2$}
		\psline[linecolor=black,linewidth=0.7pt]{->}(4.5,2.6)(4,0.4)	\rput(4.55,1.5){\footnotesize${\mathbf h}_K$}
		
		\pscircle*(4.5,2.6){0.07} 
		{\rput(4.4,2.85){\footnotesize SBS}}
		\pscircle*(6.5,3.4){0.07} 
		{\rput(6.9,3.4){\footnotesize ${\rm PU}_1$}}
		\pscircle*(6.7,2.6){0.07} 
		{\rput(7.1,2.6){\footnotesize ${\rm PU}_L$}}
		{\uput[0]{10}(6.05,3.08){\footnotesize $\vdots$}}
		\pscircle*(6.3,1.2){0.07}
		{\rput(6.68,1.2){\footnotesize ${\rm SU}_1$}}
		\pscircle*(5.5,0.6){0.07} 
		{\rput(5.85,0.6){\footnotesize ${\rm SU}_2$}}
		{\uput[0]{95}(4.15,1.0){\footnotesize $\vdots$}}
		\pscircle*(4,0.4){0.07}
		{\rput(4.4,0.4){\footnotesize ${\rm SU}_K$}}
		
		%	\pscircle*(4.0,4){0.07} \rput(4.27,4){$i$}
		%	\pscircle*(5.0,3){0.07} \rput(5.0,3.25){$1$}
		%\pscircle*(0.0,3){0.07} \rput(0.0,3.25){$-1$}
		%	\pscircle*(4.0,2){0.07} \rput(4.27,2){$-i$}
		
		%	{\color{red}\rput[bl](5.7,3.55){$y_i$}}
		%	\psline[linecolor=red,opacity=0.9,linewidth=1.7pt]{->}(4.0,3.0)(5.7,3.75)	

		%	\rput[bl](4.4,3.04){$\theta$}
		%	\rput[bl](0.0,4.5){\footnotesize${\operatorname{Im}}$}
		%	\rput[bl](3.0,1.5){\footnotesize${\operatorname{Re}}$}
		%	\rput[bl](4.0,4.5){\footnotesize${\operatorname{Im}}$}
		%	\rput[bl](7.0,1.5){\footnotesize${\operatorname{Re}}$}

		\rput[bl](1.5,0.0){($a$)}
		\rput[bl](5.5,0.0){($b$)}
		
		%	\psgrid
		\end{pspicture}
	}
	\vspace{0.1cm}
	\caption{ (a) The basic cognitive Z-channel and (b) The cognitive radio with $K$ SUs and $L$ PUs in SBS network.}
	\label{system}
\end{figure}

In line with the above, this paper extends the work on the downlink beamforming optimization problem by exploiting the constructive interference \cite{Masouros2015,Alodeh2015} to the CR Z-channel scenarios where it was previously inapplicable. Vishwanath, Jindal, and Goldsmith \cite{Vishwanath2003} introduced the Z-channel  shown in Fig.~\ref{system}(a), in which only the interference from the SBS to the PUs is considered, while the interference from the PBS to the SUs is neglected or can be modelled as additive noise at the SUs. We assume that the TDD protocol is applied, instantaneous CSI is available at the transmitter and instantaneous SU transmit data information are utilized at the SBS, as in \cite{Masouros2015,Alodeh2015}. We formulate the beamformer design problem to minimize the worst SU's symbol error probability (WSUSEP)
subject to total transmit power and PU instantaneous interference constraints, where WSUSEP is defined as the probability that worst SU wrongly decodes its symbol.
The major contributions of this paper can be summarized as follows:
\begin{enumerate}
		\item We formulate the WSUSEP beamformer design for the CR network that exploits constructive interference within the secondary link, subject to instantaneous interference constraints to the PUs.
		\item We derive conditions under which the probabilistic beamforming design allows a reformulation as a convex deterministic beamforming problem that can be efficiently solved using, e.g., the barrier method and show that these conditions are generally met in practical scenario.
		\item We derive a simple and computationally efficient approximation technique with remarkably low computational complexity in terms of average execution time that achieves close to optimal performance and allows a convenient SOCP reformulation. 
\end{enumerate}

All the above algorithms are shown to offer an improved performance-complexity trade-off compared to existing CR beamforming techniques.

{\it Remark 1:} The above generic concept of interference exploitation can be applied to a number of related CR beamforming techniques such as the CSI-robust beamformers of \cite{Wajid2013} amongst others. To constrain our focus on the proposed concept, however, here we concentrate on the CR beamforming of \cite{Fu2009,Zhang2010}, which we use as our reference and main performance benchmark. We designate the application of the constructive interference concept to alternative CR precoders as the focus of our future work.

{\it Remark 2:} In the following analysis, we consider the phase-shift keying (PSK) modulation. This is motivated by the fact that our proposed schemes are most suitable for high interference scenarios where typically low order PSK modulations are employed to secure reliable transmission. Nevertheless, by enlarging the correct detection
modulation region to exploit constructive interference, it has been shown in \cite{Masouros2013,Alodeh2015a} that the exploitation of constructive interference can be extended to other modulation schemes such as quadrature amplitude modulation (QAM).  In further work, we are looking forward to extending our proposed constructive interference-based approaches to QAM using the similar techniques given in \cite{Masouros2013,Alodeh2015a}.

The remainder of the paper is organized as follows. Section II introduces the signal model and then revisits the conventional CR downlink beamforming problem. In Section III, the constructive interference exploitation is introduced and the WSUSEP-based beamforming problem for CR networks is presented. Section IV develops an approximate approach of solving the WSUSEP-based CR downlink beamforming problem. Section V provides simulation results. Conclusions are drawn in Section VI.

\textit{Notation:}
 ${\rm E}\{\cdot\}$, ${\rm Pr}(\cdot)$, $|\cdot|$, $\|\cdot\|$, $(\cdot)^{*}$， $(\cdot)^{T}$, $\arg(\cdot)$, denote the statistical expectation, the probability function, the absolute value, the Euclidean norm, the complex conjugate, and the transpose, the angle in a complex plane between the positive real axis to the line joining the point to the origin, respectively. $\mathbf{I}_j$, and ${\mathbf 0}_{j,j}$ denotes the $j\times j$ identity matrix, and $j\times j$ zero matrix, respectively. $\mod$ is defined to be the modulo operation.
${\rm blkdiag}({\mathbf a}_1,...,{\mathbf a}_n)$ is the block diagonal matrix where ${\mathbf a}_i$ are on main diagonal blocks such that the off-diagonal blocks are zero matrices. $\operatorname{Re}(\cdot)$ and $\operatorname{Im}(\cdot)$ are the real part, and the imaginary part, respectively.
%\begin{figure*}
%\end{figure*}

\section{System Model and Conventional Downlink Beamforming Problem}
We consider a single cell CR Z-channel system, which consists of a single $N$-antenna SBS, $K$ single-antenna SUs and $L$ single-antenna PUs shown in Fig.~\ref{system}(b). The signal transmitted by the SBS is given by the $N \times 1$ vector 
\begin{equation}
{\mathbf x}=\sum_{i=1}^K{\mathbf w}_i b_i,
\label{eq:transmitted_signal}
\end{equation}
where $b_{k}\triangleq e^{j\vartheta_k}$ is the unit amplitude $M$-order PSK ($M$-PSK) modulated symbol, $\vartheta_k\triangleq k \pi/M$ is the phase of the constellation point for $k$th transmit data symbol, and ${\mathbf w}_k$ is the $N \times 1$ beamforming weight vector for the $k$th SU.  Let ${\mathbf h}_i$ be the $N \times 1$ channel vector from SBS to the $i$th SU. The received signal of the $i$th SU is 
\begin{eqnarray}
y_i \!\!\!\!&=&\!\!\!\! {\mathbf h}_i^T {\mathbf x} + n_i = %\nonumber\\&=& 
	 \underbrace{{\mathbf h}_i^T{\mathbf w}_i b_i}_\text{desired signal} + \underbrace{\sum\limits_{j=1,j\not = i}^K{\mathbf h}_i^T{\mathbf w}_jb_j + n_i}_\text{interference plus noise},
\label{eq:received_signal}
\end{eqnarray}
where $n_i$ at the $i$th SU is a circularly symmetric complex Gaussian with zero mean, i.e., $n_i \sim \mathcal{CN}(0,\sigma^2)$ and $\sigma^2$ is the noise variance for all SUs. 
The received SINR for the $i$th SU is generally expressed as the average desired signal power divided by the average interference and noise power \cite{Wajid2013}, i.e.,
\begin{eqnarray}
{\rm SINR}_i&\triangleq&\frac{{\rm E}\{|{\mathbf h}_i^T{\mathbf w}_i{b}_i|^2\}}{\sum_{\substack{j=1\\j\not = i}}^K{\rm E}\{|{\mathbf h}_i^T{\mathbf w}_j{b}_j|^2\} + {\rm E}\{|n_i|^2 \}} \nonumber\\
&=&\frac{|{\mathbf h}_i^T{\mathbf w}_i|^2}{\sum_{\substack{j=1\\j\not = i}}^K|{\mathbf h}_i^T{\mathbf w}_j|^2 + \sigma^2}.
\label{eq:SINR}
\end{eqnarray}
In \cite{Zhang2010}-\cite{Fu2009}, it is common to assume the independence of the symbols transmitted to different users, i.e., ${\rm E}\{{b}_j^*{b}_i\}=0$ for $i\not = j$, the average interference power over transmit symbols at the $l$th PU can be written as \cite{Zhang2010}
\begin{eqnarray}
{\cal I}_l
\!\!\!\!&=&\!\!\!\!{\rm E}\{\|\sum_{i=1}^K{\mathbf g}_l^T{\mathbf w}_i{b}_i\|^2\} \nonumber\\
\!\!\!\!&=&\!\!\!\!{\rm E}\{\sum_{j=1}^K\sum_{i=1}^K{b}_j^*{b}_i{{\mathbf w}_j}^H{\mathbf g}_l^*{\mathbf g}_l^T{\mathbf w}_i\}=\sum_{i=1}^K\|{\mathbf g}_l^T{\mathbf w}_i\|^2,
\label{eq:Inf}
\end{eqnarray}
where ${\mathbf g}_l$ is the $N \times 1$ channel vector between the SBS and $l$th PU. 
%Note that the above assumption is not applied in our proposed approaches. 
The average total transmitted power $P_T$ over transmit symbols is given by
\begin{eqnarray}
P_T&=&{\rm E}\{\|\sum_{i=1}^K{\mathbf w}_i{b}_i\|^2\}=\sum_{i=1}^K\|{\mathbf w}_i\|^2.
\label{eq:power}
\end{eqnarray}
In the following we present the two most common SINR-based CR downlink beamforming designs in the literature \cite{Fu2009,Zhang2010,Wajid2013}, which we will use as a reference for our proposed schemes. It is intuitive that the proposed concept can be applied to variations of these conventional beamforming problems.  
\subsection{Max-Min Fair Problem}
The conventional SINR balancing CR downlink beamforming problem aims to maximize the minimum SINR subject to average interference and total transmitted power constraints. The problem can be written as \cite{Fu2009,Zhang2010}
\begin{subequations}
	\label{eq:CR}
\begin{eqnarray}
\!\!\!\!\!\!\max_{\!{\mathbf w}_i, \gamma}  \!\!\!\!\!&&\!\!\!\!\! \gamma \nonumber\\
\text{s.t.} \!\!\!\!\!&&\!\!\!\!\! \frac{|{\mathbf h}_i^T{\mathbf w}_i|^2}{\sum_{\substack{j=1\\j\not = i}}^K|{\mathbf h}_i^T{\mathbf w}_j|^2 + \sigma^2} \geq \gamma , \; i\!=\!1,\!\dots,\!K, \label{eq:sinr}\\
\!\!\!\!\! && \!\!\!\!\! \sum_{i=1}^K\|{\mathbf w}_i\|^2 \!\leq\! P_0,  \sum_{i=1}^K\|{\mathbf g}_l^T{\mathbf w}_i\|^2 \!\leq\! \epsilon_l , \; l\!=\!1,\!\dots,\!L,	\label{eq:conventional1}
\end{eqnarray}
\end{subequations}
where $P_0$ is the total transmitted power budget and $\epsilon_l$ is the maximum admitted interference power caused by the SBS at the $l$th PU. The authors in \cite{Zhang2010} offered the fundamental approach based on the conventional downlink beamforming technique \cite{gershman2010convex}, while the authors in \cite{Fu2009} provided the most efficient implementation. Problem (\ref{eq:CR}) is feasible if the interference and power constraints in (\ref{eq:conventional1}) have a non-zero feasible point. 
By rotating the phase of ${\mathbf h}_i^T{\mathbf w}_i$, it can be assumed w.l.o.g. that ${\mathbf h}_i^T{\mathbf w}_i$ is real-valued and the solution still satisfies the constraints in (\ref{eq:sinr}). Problem (\ref{eq:CR}) can be rewritten as \cite{Fu2009,Zhang2010}
\begin{eqnarray}
\max_{\!{\mathbf w}_i, \gamma} \!\!\!\!\!&&\!\!\!\!\! \gamma \nonumber\\
\text{s.t.} \!\!\!\!\!&& \!\!\!\!\! \operatorname{Im}({\mathbf h}_i^T{\mathbf w}_i) = 0,\; i\!=\!1,\!\dots,\!K, \nonumber\\
\!\!\!\!\!&& \!\!\!\!\! \left\| \begin{array}{c}
({\mathbf I}_K \otimes {\mathbf h}_i^T) {\mathbf w}\\
\sigma
\end{array} \right\| \leq \sqrt{1 + \frac{1}{\gamma}} {\mathbf h}_i^T{\mathbf w}_i , \; i\!=\!1,\!\dots,\!K, \nonumber\\
\!\!\!\!\!&& \!\!\!\!\! \|{\mathbf w}\| \leq \sqrt{P_0},\; \|{\mathbf C}_{K+l}{\mathbf w}\| \leq \sqrt{\epsilon_l}  , \; l\!=\!1,\!\dots,\!L, \label{eq:conventional1a}
\end{eqnarray}
where $\otimes$ is a Kronecker product and ${\mathbf w}$ and ${\mathbf C}_{K+l}$ are $NK \times 1$ and $K \times NK$ such that
\begin{eqnarray}
{\mathbf w} \!\!\!\!\!&\triangleq&\!\!\!\!\!({\mathbf w}_1^T\;{\mathbf w}_2^T\dots {\mathbf w}_K^T)^T, \\
{\mathbf C}_{K+l}  \!\!\!\!\!&\triangleq&\!\!\!\!\! {\rm blkdiag}(\underbrace{{\mathbf g}_l^T,{\mathbf g}_l^T,...{\mathbf g}_l^T}_\text{$K$ times}).
\end{eqnarray}
Problem (\ref{eq:conventional1a}) is a quasi-convex optimization problem and can be solved using the bisection method and sequential SOCP. 
Nevertheless, the above problem does not take instantaneous interference exploitation into account for the transmit data symbols as a part of the optimization problem for each transmission. Moreover, our results in the simulations show that despite the average interference constraints in (\ref{eq:conventional1a}), the instantaneous interference may violate the interference power constraints, leading to outages for the PUs.

In the next section, we consider to design the CR downlink beamforming problem by making use of the instantaneous transmit data symbols to exploit constructive interference within the secondary links and restrict the instantaneous interference created by the SBS within primary links. 
\begin{figure*}
	%\begin{figure}[t]
	\centering
	\psscalebox{1.15}{
		\begin{pspicture}(16,5)
		
		\psset{blendmode=2}
		
		\psset{linewidth=1.0pt}
		
		\def\radius{1.0}       % Radius of L1-Ball
		
		\def\Theta{60 } \def\Radius{4 }
		
		% Axis
	%	\psset{arrowsize=3.5pt,arrowinset=0.02}
	%	\psline[linewidth=0.7pt]{->}(0,1.5)(3,1.5)
	%	\psline[linewidth=0.7pt]{->}(0,0.0)(0,4.5)
		
	%	\psline[linecolor=black,linewidth=0.7pt,linestyle=dashed]{-}(0.0,1.5)(3,4.5)
		
	%	\pspolygon*[linecolor=black!80!white,opacity=0.3](0.4,1.9)(1.4,4.9)(3.4,2.9)
	%	\pspolygon*[linecolor=black!80!white,opacity=0.3](0,1.5)(1,4.5)(3,2.5)
		
	%	\pscircle*(0.707,2.207){0.07} 	
		
	%	\psline[linecolor=black,linewidth=0.7pt]{|<->|}(0.0,1.5)(0.4,1.9)\uput[-90]{45}(0.45,1.85){\footnotesize$\tau\sigma$}
		
	%	\psline[linecolor=blue,linewidth=0.7pt]{|<->|}(0,1.5)(1.574,3.074)\uput[-90]{45}(1.7,3.2){\footnotesize ${\operatorname{Re}(b_i^*{\mathbf h}_i^T {\mathbf x})}$}
	%	\psline[linecolor=blue,linewidth=0.7pt]{|<->|}(1.574,3.074)(1.15,3.5)\uput[0]{45}(1.15,3.5){\footnotesize ${\operatorname{Im}(b_i^*{\mathbf h}_i^T {\mathbf x})}$}

	%	{\color{blue}\uput[90]{45}(0.8,2.15){\footnotesize$\phi_i$}}
	%	\psline[linecolor=blue,linewidth=1.0pt]{->}(0.4,1.9)(1.15,3.5)
	%	\psarc[linecolor=blue,linewidth=1.0pt](0.4,1.9){0.95}{45}{65.1}

	%	{\color{red}\uput[90]{45}(0.8,3.2){\footnotesize${\mathbf h}_i^T {\mathbf x}$}} 
	%	\psline[linecolor=red,opacity=0.9,linewidth=1.0pt]{->}(0,1.5)(1.15,3.5)

	%	\psarc[linewidth=1.0pt](0.4,1.9){1.35}{45}{71.565}
	%	\uput[90]{45}(1.15,2.5){\footnotesize$\theta$}

	%	\psline[linecolor=black,linewidth=0.7pt]{->}(2.4,2.9)(2.8,3.3)

		% Axis
		\psset{arrowsize=3.5pt,arrowinset=0.02}
		\psline[linewidth=0.7pt]{->}(1,1.5)(4,1.5)
		\psline[linewidth=0.7pt]{->}(1,0.0)(1,4.5)
		
		\pspolygon*[linecolor=black!80!white,opacity=0.3](1,1.5)(2,4.5)(4,2.5)
		\pscircle*(1.707,2.207){0.07} \uput[0]{45}(1.6,2.1){\footnotesize$b_i$}		
		
		\psline[linecolor=black,linewidth=0.7pt,linestyle=dashed]{-}(1.0,1.5)(4,4.5)

		{\color{red}\uput[90]{45}(2,3.4){\footnotesize$y_i$} \uput[90]{45}(1.9,2.25){\footnotesize$\psi_i$}}
		\psline[linecolor=red,opacity=0.9,linewidth=1pt]{->}(1.0,1.5)(2,3.5)
		\psarc[linecolor=red,linewidth=1.0pt](1.0,1.5){1.2}{45}{63.435}
		
		\psarc[linewidth=1.0pt](1.0,1.5){0.8}{45}{71.565}
		\uput[90]{45}(1.35,1.7){\footnotesize$\theta$}

		\psarc[linewidth=0.7pt]{<-}(5,1.5){2}{20}{100}
		\rput[bl](6.2,3.2){\footnotesize$\angle b_i^*$}
		
		% Axis
		\psset{arrowsize=3.5pt,arrowinset=0.02}
		\psline[linewidth=0.7pt]{->}(6.0,1.5)(9,1.5)
		\psline[linewidth=0.7pt]{->}(6,0.0)(6,4.5)
		
		\pspolygon*[linecolor=black!80!white,opacity=0.3](6,1.5)(8.828,2.914)(8.828,0.186)
		\pscircle*(7,1.5){0.07} \rput[bl](6.93,1.1){\footnotesize$1$}

		{\color{red}\rput[bl](8.121,2.207){\footnotesize$y_ib_i^*$} \rput[bk](7.45,1.52){\footnotesize$\psi_i$}}
		\psline[linecolor=red,opacity=0.9,linewidth=1pt]{->}(6.0,1.5)(8.121,2.207)
		\psarc[linecolor=red,linewidth=1.0pt](6.0,1.5){1.2}{0}{18.435}
		
		\psline[linecolor=blue,linewidth=1pt]{->}(6.0,1.5)(8.121,1.5)
		\rput[lb](8.2,1.8){\footnotesize ${\operatorname{Im}(y_ib_i^*)}$}
		\psline[linecolor=blue,linewidth=1pt]{<-}(8.121,2.207)(8.121,1.5)
		\rput[lb](7.4,1){\footnotesize ${\operatorname{Re}(y_ib_i^*)}$}

		\psarc[linewidth=1.0pt](6.0,1.5){0.8}{0}{26.565}
		\rput[bl](6.5,1.52){\footnotesize$\theta$}

		% Axis
		\psset{arrowsize=3.5pt,arrowinset=0.02}
		\psline[linewidth=0.7pt]{->}(11,1.5)(14,1.5)
		\psline[linewidth=0.7pt]{->}(11,0.0)(11,4.5)	
		
		\rput*(13.25,2.3){\footnotesize$\Upsilon\!\sigma$} 
		
		\pspolygon*[linecolor=black!80!white,opacity=0.3](11,1.5)(13.828,2.914)(13.828,0.186)
		\pscircle*(12,1.5){0.07} \rput[bl](11.93,1.1){\footnotesize$1$}

		{\color{red}\rput[bl](12.8,1.8){\footnotesize$y_ib_i^*$}}
		\psline[linecolor=red,opacity=0.9,linewidth=1pt]{->}(11,1.5)(13.121,2.207)
		
		\psline[linecolor=blue,linewidth=1pt]{->}(11,1.5)(13.5,1.5)
		\rput[lb](13.6,1.8){\footnotesize ${\operatorname{Im}(b_i^*{\mathbf h}_i^T {\mathbf x})}$}
		\psline[linecolor=blue,linewidth=1pt]{<-}(13.5,2.207)(13.5,1.5)
		\rput[lb](12.6,1){\footnotesize ${\operatorname{Re}(b_i^*{\mathbf h}_i^T {\mathbf x})}$}

		\pscircle*[linecolor=red,opacity=0.3](13.5,2.207){0.5}

		\psline[linecolor=blue,linewidth=1pt]{|<->|}(13.5,2.207)(13.5,2.75)
		\psline[linewidth=0.5pt]{|-|}(13.3,2.65)(13.5,2.207)
		\rput[lb](13.6,2.3){\footnotesize $\frac{\Upsilon\sigma }{\cos \theta} $}
		\psarc[linewidth=0.5pt](13.5,2.207){0.3}{90}{116.565}

		\psarc[linewidth=1.0pt](11,1.5){0.8}{0}{26.565}
		\rput[bl](11.5,1.52){\footnotesize$\theta$}

	%	\rput[bl](0.0,4.5){\footnotesize${\operatorname{Im}}$}
	%	\rput[bl](3.0,1.5){\footnotesize${\operatorname{Re}}$}
		\rput[bl](1.0,4.5){\footnotesize${\operatorname{Im}}$}
		\rput[bl](4.0,1.5){\footnotesize${\operatorname{Re}}$}
		\rput[bl](6.0,4.5){\footnotesize${\operatorname{Im}}$}
		\rput[bl](9.0,1.5){\footnotesize${\operatorname{Re}}$}
		\rput[bl](11,4.5){\footnotesize${\operatorname{Im}}$}
		\rput[bl](14,1.5){\footnotesize${\operatorname{Re}}$}
		
	%	\rput[bl](7,2.5){\footnotesize$\rm {I}$}
	%	\rput[bl](5,2.5){\footnotesize$\rm {II}$}
	%	\rput[bl](5,0.5){\footnotesize$\rm {III}$}
	%	\rput[bl](7,0.5){\footnotesize$\rm {IV}$}

	%	\rput[bl](1.5,0.0){($a$)}
		\rput[bl](2.5,0.0){($a$)}
		\rput[bl](7.5,0.0){($b$)}
		\rput[bl](12.5,0.0){($c$)}
		%	\psgrid
		\end{pspicture}
	}
	\vspace{0.1cm}
	\caption{For $M$-PSK modulation, (a) constructive interference $y_i$ within correct detection region where the constructive area of constellation is indicted by the grey area; (b) after rotation by $\angle b_i^*$, ${\operatorname{Re}(y_ib_i^*)}$ and $\operatorname{Im}(y_ib_i^*)$ are projected from $y_ib_i^*$ on real and imaginary axis, respectively; (c) constructive interference is described using trigonometry. %(a) precoding for interference-based optimization \cite{Masouros2015}; 
	}
	\label{constellation}
	%\end{figure}
\end{figure*}
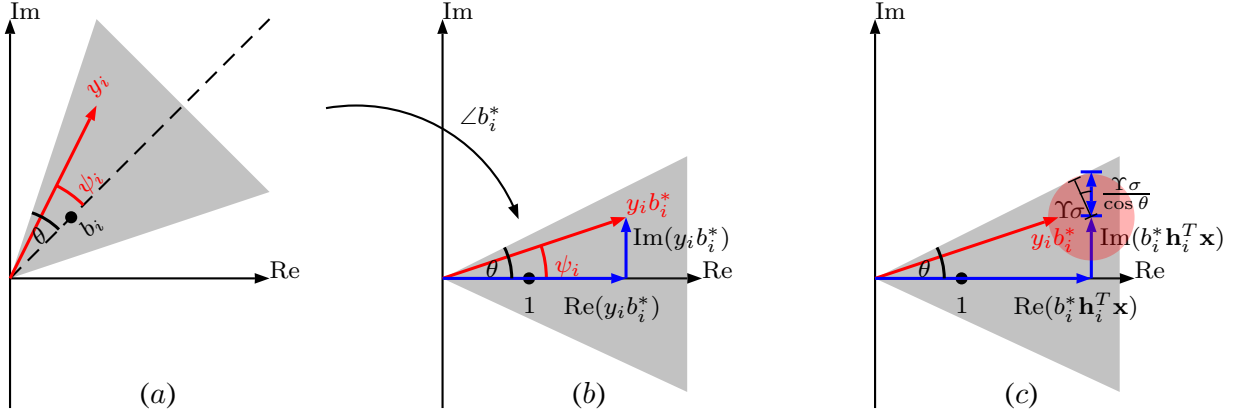

\section{WSUSEP-based CR downlink beamforming for interference exploitation} \label{WorstSERsection}
\subsection{Constructive interference Exploitation} %(\ref{eq:constructiveInterference}) and (\ref{eq:constructiveInterference1}) we derive CR beamforming optimization by introducing a noise robust adaptation. first 
In the conventional downlink beamforming problems \cite{Rashid1998,schubert2004solution,bengtsson1999optimal, bengtsson2001optimal, Shamai2006, gershman2010convex, Choi2015},  beamformers are designed by mitigating the average mutiuser interference, in which suppressing interference requires some degrees of freedom. It has been established in \cite{Masouros2015} that given the instantaneous transmit data symbols and CSI at the transmitter, it is not necessarily required to completely suppress the interference; instead the beamformers can be designed to constructively use the interfering signal to enhance the desired signal. With the aid of exploiting the instantaneous interference and adapting the beamformers, the constructive interference can alter the received signals further into the correct detection region to improve the system performance. Inspired by this idea, we provide a systematic treatment of constructive interference as illustrated in Fig.~\ref{constellation}(a), where the nominal PSK constellation point is represented by the black circle. According to \cite{Masouros2015}, we say that the received signal $y_i$ exploits the interference constructively if $y_i$ falls within the correct detection region, which is the shaded area shown in Fig.~\ref{constellation}(a). Let $\psi_i$ in Fig.~\ref{constellation}(a) denote the angle between the received signal $y_i$ and the transmitted symbol $b_i$ in the complex plane. According to (\ref{eq:received_signal}), the angle $\psi_i$ depends on the transmitted signal $\mathbf{x}$ and the noise $n_i$. Hence the angle $\psi_i$ can be treated as a function of $\mathbf{x}$ and $n_i$, i.e.,  
\begin{eqnarray}
\psi_i(\mathbf{x},n_i)&=& (\arg y_i - \arg b_i) \mod 2 \pi \nonumber\\
&=& \arg(y_i b_i^*) =\tan^{-1}\Bigl(\frac{\operatorname{Im}(y_ib_i^*)}{\operatorname{Re}(y_ib_i^*)}\Bigr), %&  {\operatorname{Re}(y_ib_i^*)} \geq 0
\label{psiangle}
\end{eqnarray}			
where ${\operatorname{Im}(y_ib_i^*)}$ and ${\operatorname{Re}(y_ib_i^*)}$ are the projections of $y_ib_i^*$ onto the real and imaginary axis, respectively. 
The product $y_i b_i^*$ is displayed in Fig.~\ref{constellation}(b) along with the corresponding decision region and the angle $\psi_i(\mathbf{x},n_i)$. The received signal $y_i$ of the $i$-th user is detected correctly, if and only if 
\begin{eqnarray}
	\psi_i(\mathbf{x},n_i) &\in& {\cal A}_{-\theta}^{\theta}, \quad\quad i=1,\dots,K,
	\label{eq:classification}
	\end{eqnarray}
	where the angular set
	\begin{eqnarray}
	{\cal A}_{\theta_1}^{\theta_2} \triangleq \bigl\{ \tilde\psi \!\!\! \mod 2 \pi \mid \theta_1 \leq \tilde \psi \leq  \theta_2, \tilde \psi\in \mathbb{R} \bigr\},
	\label{eq:set}
	\end{eqnarray}
	defines the decision region and $\theta=\pi/M$ is the maximum angular shift for an M-PSK constellation. 
Detailed deviations of the constructive interference regions for generic PSK modulations can be found in \cite{Masouros2015} and references therein.
	Based on above definition and discussion, we formulate in the following section the CR beamformer design to exploit the instantaneous interference.

\subsection{WSUSEP Approach} \label{WorstSERsection}
In this section, we derive the WSUSEP-based CR downlink beamforming problem.\footnote{In this paper, we do not consider the power minimization problem as the power minimization solutions can be derived by the corresponding solutions to the WSUSEP optimization. } The idea of this approach is to design the beamformers to steer the receive signals of SUs into the corresponding decision regions to reduce the corresponding symbol error. Furthermore, since the distribution of noise is known, we can calculate the symbol error probability (SEP) for each SU and use the WSUSEP as an objective function. The beamformer design minimizes the WSUSEP subject to the instantaneous total transmit power and instantaneous  interference power constraints, which can be written as 
\begin{subequations}	
\label{eq:worstser}
\begin{eqnarray}
\!\!\!\!\!\!\min_{\!\mathbf{x}, \rho}  \!\!\!\!&&\!\!\!\! \rho \nonumber\\
\text{s.t.} \!\!\!\!&&\!\!\!\! 
\operatorname{Pr}\bigl( {\psi_i}(\mathbf{x},n_i) \in {\cal A}_{\theta }^{2\pi - \theta} \bigr) \leq \rho, \; i=1,\dots,K, \label{probconst1}\\
\!\!\!\!&&\!\!\!\! \|\mathbf{x}\|^2 \leq P, \|{\mathbf g}_{l}^T{\mathbf x}\|^2 \leq \epsilon_l , \; l=1,\dots,L,\label{rhoconstrain}	 	
\end{eqnarray}
\end{subequations}
where $\rho$ models the WSUSEP, $\operatorname{Pr}\bigl( {\psi_i}(\mathbf{x},n_i) \in {\cal A}_{\theta }^{2\pi - \theta} \bigr)$
is $i$th SU's SEP, i.e., the probability that the received signal falls outside the correct detection region and ${\psi_i}(\mathbf{x},n_i) \notin {\cal A}_{-\theta}^{\theta}$,  
$\|{\mathbf x}\|^2$ is the instantaneous total transmitted power from the SBS, and
$\|{\mathbf g}_{l}^T{\mathbf x}\|^2$ is the instantaneous interference power for SBS to the $l$th PU.
%, $\rho_0$ in (\ref{rhoconstrain}) is the predefined maximal SEP threshold for all SUs to ensure acceptable SEPs. 
By considering the complement of the symbol error set, 
(\ref{probconst1}) can be reformulated as 
\begin{eqnarray}
1 -\operatorname{Pr}\bigl( {\psi_i}(\mathbf{x},n_i) \in {\cal A}_{-\theta}^{\theta} \bigr) \leq\rho. \label{probconst2}
\end{eqnarray}
First let us simplify the set ${\cal A}_{-\theta}^{\theta}$ in (\ref{probconst2}), i.e., (\ref{eq:classification}). By (\ref{psiangle}), the classification criteria (\ref{eq:classification}) can be directly reformulated as the following alternatives
\begin{subequations}
\begin{eqnarray}
{\rm I:} &\frac{|\operatorname{Im}(y_ib_i^*)|}{\operatorname{Re}(y_ib_i^*)} \leq \tan \theta, \;\; &\mbox{ for }\; {\operatorname{Re}(y_ib_i^*)} > 0,\\
%\mbox{or }&&\nonumber\\
{\rm II:}&y_ib_i^*= 0,  \;\; &\mbox{ for }\; {\operatorname{Re}(y_ib_i^*)} = 0, \label{eq:real_contraints}
\end{eqnarray}
\end{subequations}
 which is equivalent to the single inequality	
\begin{equation}
{|\operatorname{Im}(y_ib_i^*)|} - {\operatorname{Re}(y_ib_i^*)}\tan \theta \leq 0.
\label{eq:combine_contraints}
\end{equation}
In this paper, we only consider M-PSK modulation schemes with $M \geq 4$.\footnote{Note that $\tan \theta=\infty$ for $M=2$ as $\theta=\pi/2$. In this case, the constraint in (\ref{probconst2}) can be formulated as $1 \!-\!\operatorname{Pr}( {\operatorname{Re}(y_ib_i^*)} \geq 0 ) \!\leq\!\rho$, which can reformulated using the univariate normal cumulative distribution function. The corresponding optimization in (\ref{eq:worstser}) is convex when $\rho^{\star} \leq 0.5$ where $\rho^{\star}$ is the optimal value of (\ref{eq:worstser1}), i.e., ${\operatorname{Re}(b_i^*{\mathbf h}_i^T {\mathbf x})} \geq 0$.} 
Introducing the real-valued parameter vector representation
\begin{eqnarray}
\bar  {\mathbf x}&\triangleq&[\operatorname{Re}({\mathbf x})^T,\; \operatorname{Im}({\mathbf x})^T ]^T, \label{eq:tildex}\\
\bar {\mathbf h}_i&\triangleq&[\operatorname{Im}(b_i^*{\mathbf h}_i)^T, \; \operatorname{Re}(b_i^*{\mathbf h}_i)^T]^T,
\end{eqnarray}
we can express the real and imaginary part of the transmitted signal in (\ref{eq:combine_contraints}) as follows
\begin{eqnarray}
\operatorname{Re}(b_i^*{\mathbf h}_i^T {\mathbf x})&=&\bar {\mathbf h}_i^T {\mathbf \Pi}_K \bar  {\mathbf x}, \label{eq:real}\\
\operatorname{Im}(b_i^*{\mathbf h}_i^T {\mathbf x})&=&\bar {\mathbf h}_i^T \bar  {\mathbf x},  \label{eq:imagine}
\end{eqnarray}
where ${\mathbf \Pi}_K\triangleq[{\mathbf 0}_{K,K} \; -{\mathbf I}_K; {\mathbf I}_K \; {\mathbf 0}_{K,K}]$ is a selection matrix.
Resolving the absolute value term in (\ref{eq:combine_contraints}), we obtain two linear inequalities
\begin{subequations}
\begin{eqnarray}
&{\mathbf t}_{2i-1}^T\bar  {\mathbf x} \!\!\!\!&\geq \tilde n_{2i-1}, \\
&{\mathbf t}_{2i}^T\bar  {\mathbf x} \!\!\!\!&\geq \tilde n_{2i}, 
\end{eqnarray}
\label{eq:contraints2}
\end{subequations}
where
\begin{eqnarray}
&{\mathbf t}_{2i-1}^T \!\!\!\!& \triangleq-\bar {\mathbf h}_i^T + \tan \theta \;\bar {\mathbf h}_i^T {\mathbf \Pi}_K,\\
&{\mathbf t}_{2i}^T \!\!\!\!& \triangleq \bar {\mathbf h}_i^T + \tan \theta \; \bar {\mathbf h}_i^T {\mathbf \Pi}_K, \\
 &\tilde n_{2i-1} \!\!\!\!& \triangleq \operatorname{Im}(b_i^*n_i) -  {\operatorname{Re}(b_i^*n_i)}\tan \theta, \label{equni}\\
&\tilde n_{2i} \!\!\!\!& \triangleq -\operatorname{Im}(b_i^*n_i) -  {\operatorname{Re}(b_i^*n_i)}\tan \theta. \label{equni2}
\end{eqnarray}
The vectors ${\mathbf t}_{j}, \; j=1,...,2K,$ are deterministic and depend on the channel and the decision region defined by the angle $\theta$, and the scalars $\tilde n_{j}$ are real-valued Gaussian random variables (linear transformations of Gaussian random variables). 
By (\ref{eq:contraints2}), the probability function in (\ref{probconst2}) can be written as a joint probability function
\begin{eqnarray}
\operatorname{Pr}\Bigl(\! {\mathbf t}_{2i-1}^T\bar {\mathbf x}  \!\geq\! \tilde n_{2i-1},{\mathbf t}_{2i}^T\bar {\mathbf x}  \!\geq\! \tilde n_{2i} \!\Bigr). \label{probc1}
\end{eqnarray}
Consider the bivariate standard normal probability distribution $\phi$ with zero mean such that
\begin{eqnarray}
\phi({\mathbf u};r)=\frac{1}{2\pi \sqrt{1- r^2}}\exp(-\frac{1}{2}{\mathbf u}^T{\boldsymbol{\Sigma}}^{-1}{\mathbf u}),
\end{eqnarray}
where ${\mathbf u}\triangleq[u_1,\;u_2]^T$, the correlation $r$ is defined as
\begin{eqnarray}
r\triangleq\frac{{\rm E}\{\eta_1 \eta_2\}}{\sqrt{{\rm E}\{|\eta_1|^2\}{\rm E}\{|\eta_2|^2\}}}={\rm E}\{\eta_1 \eta_2\},
\end{eqnarray}
with $|r|<1$,  $\eta_1,\eta_2$ are the standardized random variables, i.e., ${\rm E}\{|\eta_1|^2\}={\rm E}\{|\eta_2|^2\}=1$, and
\begin{eqnarray}
{\boldsymbol{\Sigma}}\triangleq {\rm E}\bigl\{ [\eta_1, \eta_2]^T[\eta_1, \eta_2] \bigr\}=  \begin{bmatrix}1&r\\r&1 \end{bmatrix}.
\end{eqnarray}
The cumulative distribution function (CDF) of the standard bivariate normal distribution is defined by
\begin{eqnarray}
\Phi({\mathbf u};r)=\int _{-\infty}^{u_1}\int _{-\infty}^{u_2}\phi(\tilde {\mathbf u};r)\,d \tilde u_1 d \tilde  u_2.
\end{eqnarray}
Then the corresponding probability function for $u_1 \geq \eta_1,u_2 \geq \eta_2$ is given by
\begin{eqnarray}
\operatorname{Pr}\Bigl( u_1 \geq \eta_1,u_2 \geq \eta_2 \Bigr) = \Phi({\mathbf u};r). \label{equCDF}
\end{eqnarray}
Since $n_i$ in (\ref{eq:received_signal}) is a circularly symmetric zero mean complex Gaussian random variable, we can conclude that $\tilde n_j$ in (\ref{equni}) and (\ref{equni2}) is also a real-valued Gaussian with zero mean and variance
\begin{equation}
\sigma_{\tilde n}^2\triangleq\operatorname{E}\{\tilde n_j^2\}=\frac{(1 +\tan^2 \theta)\sigma^2}{2}=\frac{\sigma^2}{2 \cos^2 \theta}, \label{tildenoise}
\end{equation}
i.e., $\tilde n_j \sim \mathcal{N}(0,\frac{\sigma^2}{2 \cos \theta^2})$.
Since $\tilde n_{2i-1}$and $\tilde n_{2i}$ correspond to a real bivariate normal distribution and according to (\ref{equCDF}), we can express (\ref{probc1}) as a joint normal CDF
\begin{eqnarray}
\Phi\biggl(\biggl[\frac{{\mathbf t}_{2i-1}^T\bar {\mathbf x}}{\sigma_{\tilde n}}, \;\frac{{\mathbf t}_{2i}^T\bar {\mathbf x}}{\sigma_{\tilde n}}\biggr]^T; \bar r\biggr), \label{bicdf}
\end{eqnarray}
with the correlation of $\tilde n_{2i-1}$ and $\tilde n_{2i}$ is given by
\begin{eqnarray}
\bar r&=&\frac{-\!1 \!+\! \tan^2 \theta}{1 \!+\! \tan^2 \theta}=-\cos 2\theta. \label{correlation} 
\end{eqnarray}
By (\ref{probconst2}) and (\ref{bicdf}), problem (\ref {eq:worstser}) can be reformulated as 
\begin{subequations}
	\label{eq:worstser1}
\begin{eqnarray}
\!\!\!\!\!\!\!\min_{\!\mathbf{x}, \rho}  \!\!\!\!\!&&\!\!\!\!\! \rho \nonumber\\
\!\!\!\!\!\!\!\text{s.t.} \!\!\!\!\!&& \!\!\!\!\! 
1 \!-\!\Phi\biggl(\biggl[\frac{{\mathbf t}_{2i-1}^T\bar {\mathbf x}}{\sigma_{\tilde n}}, \;\frac{{\mathbf t}_{2i}^T\bar {\mathbf x}}{\sigma_{\tilde n}}\biggr]^T\!\!; \bar r\biggr)\!-\!\rho \!\leq\! 0,\;
 i\!=\!1,\!\dots,\!K, \label{probconstrain}\\	
\!\!\!\!\!&& \!\!\!\!\! \|\bar{\mathbf x}\| \!-\! \sqrt{P} \leq 0,%\label{feasible1}\\
%\!\!\!\!\!&& \!\!\!\!\! 
\|{\mathbf B}_{l}{\bar  {\mathbf x}}\| \!-\! \sqrt{\epsilon_l} \leq 0 , \; l=1,\dots,L,	\label{feasible2}%\\
%\!\!\!\!\!&& \!\!\!\!\!  -\rho \leq 0, \label{feasible3}	\\
%\!\!\!\!\!&& \!\!\!\!\!  \rho \!-\! \rho_0\!\leq\! 0, \label{rhoconstrain1}
\end{eqnarray}
\end{subequations}
where 
\begin{eqnarray}
{\mathbf B}_{l}&\triangleq&
\left[ \begin{array}{cc}
{\operatorname{Re}({\mathbf g}_l^T)} & -{\operatorname{Im}({\mathbf g}_l^T)}\\
{\operatorname{Im}({\mathbf g}_l^T)} & {\operatorname{Re}({\mathbf g}_l^T)}\end{array} \right]
\end{eqnarray}
is a $2 \times 2N$ real matrix.
We remark that constraint (\ref{probconstrain}) is generally non-convex. Note that the sufficient condition for the concavity of the standard bivariate normal CDF is non-trivial.  
Author in \cite{Prekopa1970} showed that $\Phi({\mathbf u}; r)$ is concave in one variable under a certain condition on $u_1$ and $u_2$, respectively. 
\newtheorem*{Lemma1}{Lemma 1A}  
\begin{Lemma1} \cite{Prekopa1970} (Concavity in one variable - positive correlation)
	Let $r \geq 0$. Then $\Phi({\mathbf u}; r)$ is concave in $u_i$ for fixed $u_j$ with $j \neq i$, i.e., $\frac{\partial^2\Phi({\mathbf u}; r)}{\partial u_i^2} \leq 0$  for $i=1,2$.
\end{Lemma1}
\newtheorem*{Lemma2}{Lemma 1B}  
\begin{Lemma2} \cite{Prekopa1970} (Concavity in one variable - negative correlation)
	Let $-1 \leq r \leq 0$. Then $\Phi({\mathbf u}; r)$ is concave in $u_i$ for fixed $u_j$ with $j \neq i$, i.e., $\frac{\partial^2\Phi({\mathbf u}; r)}{\partial u_i^2} \leq 0$  for $i=1,2$, if 
	\begin{eqnarray}
	u_i\geq \sqrt{\frac{\phi(1)}{2\Phi(1) + \phi(1)}},\; i=1,2,
	\end{eqnarray}
	where the probability density function and CDF of a standard univariate normal distribution are given by
	\begin{eqnarray}
	\phi(u)&=&\frac{1}{\sqrt{2\pi}}\exp(-\frac{{u}^2}{2}), \Phi(u)=\int _{-\infty}^{u}\phi(\tilde u)\,d \tilde u, %&=&
	\end{eqnarray}
	respectively.
\end{Lemma2}
In this paper, we further restrict the conditions on variables to guarantee the joint concavity of the CDF in (\ref{probconstrain}) and show that these conditions are generally met in conventional transmission scenarios. %The following theorem provides a condition to ensure (\ref{probconstrain}) to be a convex constraint.  
\newtheorem*{Theorem1}{Theorem 1} 
\begin{Theorem1} (Joint Concavity)
	For $M \geq 4$, the standard bivariate normal CDF in (\ref{probconstrain}) is concave if ${\mathbf t}_{j}^T\bar {\mathbf x}$ satisfies the inequality 
	\begin{eqnarray} 
	{\mathbf t}_{j}^T\bar {\mathbf x}/\sigma_{\tilde n}  \geq  \alpha^{\star}(\bar r),\; j=1,...,2K, \label{p1}
	\end{eqnarray}	
	with threshold $\alpha^{\star}(\cdot)$ denoting the optimal function value of the following constrained optimization problem: 
	  \begin{eqnarray}
	  \!\!\!\!\!\!\!\!\!\!\alpha^{\star}(r): \; \min_{\alpha} \!\!\!\!\!&& \!\!\!\!\!  \alpha  \;\text{\rm s.t.}\;   \frac{\Phi\Bigl(\alpha \frac{1 - r }{\sqrt{1 - r^2}}\Bigr) }{\phi\Bigl(\alpha \frac{1 - r }{\sqrt{1 - r^2}}\Bigr)}\alpha \geq \frac{1 - r }{\sqrt{1 - r^2}}. \label{alphaopt}
	  \end{eqnarray}
\end{Theorem1}
\begin{IEEEproof}
	See Appendix \ref{proofLMZ1}.	
\end{IEEEproof}
%\newtheorem*{Theorem4}{Theorem 4} 
%\begin{Theorem4}
%	For $M \geq 4$, the set of points satisfying the inequality 
%	\begin{eqnarray}
%	\operatorname{Pr}\Bigl( {\mathbf t}_{2i-1}^T\bar {\mathbf x}  \geq \tilde n_{2i-1},{\mathbf t}_{2i}^T\bar {\mathbf x}  \geq \tilde n_{2i} \Bigr) \geq p.\label{prob2}
%	\end{eqnarray}
%	is convex, i.e., $p - \operatorname{Pr}\Bigl( {\mathbf t}_{2i-1}^T\bar {\mathbf x}  \geq \tilde n_{2i-1},{\mathbf t}_{2i}^T\bar {\mathbf x}  \geq \tilde n_{2i} \Bigr)$ is a convex function, where $p$ is a fixed probability satisfying the inequality 
%	\begin{eqnarray} 
%	p \geq \Phi\Biggl(\sqrt{\max_{v \geq 0} \frac{\phi(v)v}{\Phi(v)}}\Biggr) \!\approx\! 0.7063. \label{p2}
%	\end{eqnarray}	
%\end{Theorem4}
%\begin{IEEEproof}
%	See Appendix.	
%\end{IEEEproof}
\noindent 
Following from (\ref{probconstrain}), we have 
\begin{eqnarray}
\Phi\biggl(\biggl[\frac{{\mathbf t}_{2i-1}^T\bar {\mathbf x}}{\sigma_{\tilde n}}, \;\frac{{\mathbf t}_{2i}^T\bar {\mathbf x}}{\sigma_{\tilde n}}\biggr]^T; \bar r\biggr) \!\geq\! 1\!-\! \rho \!\geq\! 1\!-\! \rho^{\star}, \label{phiineq}
\end{eqnarray}
where $\rho^{\star}$ is the optimal value of (\ref{eq:worstser1}).
Moreover, by the definition of univariate and bivariate normal CDFs and with $\Phi({\mathbf u}; r)$ being an increasing function on $r$ for fixed ${\mathbf u}$, we obtain
\begin{eqnarray}
\Phi\biggl(\frac{{\mathbf t}_{j}^T\bar {\mathbf x}}{\sigma_{\tilde n}}\biggr) \!\!\!\!&=&\!\!\!\! \int _{-\infty}^{\infty}\int _{-\infty}^{\frac{{\mathbf t}_{j}^T\bar {\mathbf x}}{\sigma_{\tilde n}}}\phi(\tilde {\mathbf u}; 0)\,d \tilde u_1 d \tilde u_2 \nonumber\\
\!\!\!\!&\geq&\!\!\!\! \Phi\biggl(\biggl[\frac{{\mathbf t}_{2i-1}^T\bar {\mathbf x}}{\sigma_{\tilde n}}, \;\frac{{\mathbf t}_{2i}^T\bar {\mathbf x}}{\sigma_{\tilde n}}\biggr]^T; 0\biggr) \nonumber\\
\!\!\!\!&\geq&\!\!\!\! \Phi\biggl(\biggl[\frac{{\mathbf t}_{2i-1}^T\bar {\mathbf x}}{\sigma_{\tilde n}}, \;\frac{{\mathbf t}_{2i}^T\bar {\mathbf x}}{\sigma_{\tilde n}}\biggr]^T; \bar r\biggr), \label{phiineq2}
\end{eqnarray}
for $j=2i\!-\!1,2i$. Hence, this yields
\begin{eqnarray}
\Phi\biggl(\frac{{\mathbf t}_{j}^T\bar {\mathbf x}}{\sigma_{\tilde n}}\biggr)  \!\geq\! 1\!-\! \rho^{\star}, \label{phiineq3}
\end{eqnarray}
for $j=1,...,2K$. If we assume that 
\begin{eqnarray}
1 -\rho^{\star} \geq  \Phi(\alpha^{\star}(\bar r)), \label{rhoineq}
\end{eqnarray}
then, by inequalities (\ref{phiineq3})-(\ref{rhoineq}), and the strict monotonicity property of the standard univariate normal CDF, we ensure that condition (\ref{p1}) is satisfied. Thus, by Theorem $1$, the assumption in (\ref{rhoineq}) can guarantee problem (\ref{eq:worstser1}) to be convex. That is, as of Theorem 1, for the optimal value $\rho^{\star}$ of (\ref{eq:worstser1}) such that
$1 -\Phi(\alpha^{\star}(\bar r)) \geq  \rho^{\star}$ for a given correlation $\bar r$, the optimization problem in (\ref{eq:worstser1}) is convex.
In Table~\ref{tab10}, we list, as examples, the lower bounds of (\ref{p1}) and the upper bounds of $\rho^{\star}$ for different values $M$ of the constellation size.
For example, when $M=4$, the value of $\rho^{\star}$ in (\ref{rhoineq}) corresponds to a SEP of less than $30.64\%$ which does not put any restrictions on our beamformer design as in typical applications much lower SEP values are required. Accordingly, the optimization problem in (\ref{eq:worstser1}) is convex for all practical SEP constraints. 
\begin{table}[t]
	\begin{center}
		\begin{tabular}{|r|r|r|r|}
			\hline
		$M$&$\bar r$&$\alpha^{\star}(\bar r)$&$1 - \Phi(\alpha^{\star}(\bar r))$\\
			\hline
			\hline
			$4$&$0$&$0.5061$&$0.3064$ \\
			\hline
			$8$&$-0.7071$&$0.5088$&$0.3055$ \\
			\hline
			$16$&$-0.9239$&$0.3694$&$0.3559$ \\
			\hline
			$32$&$-0.9808$&$0.2353$&$0.4070$ \\
			\hline
			$64$&$-0.9952$&$0.1400$&$0.4443$ \\
			\hline
			\hline
		\end{tabular} 
	\end{center}
	\caption{The correlation $\bar r$, lower bounds of (\ref{p1}) and upper bounds of the optimal value $\rho^{\star}$ of (\ref{eq:worstser1}) for different constellation size $M$}  
	\label {tab10}
\end{table}

Suppose the conditions in (\ref{p1}) of Theorem $1$ are satisfied, then (\ref{eq:worstser1}) can be written as a convex optiminization problem and can be solved by any contemporary methods such as the subgradient projection and barrier methods \cite{boyd2004convex}. For the sake of illustration, we choose to use the barrier method to solve (\ref{eq:worstser1}). 
%According to \cite{Prekopa1993}, we can solve (\ref{eq:worstser1}) via barrier method. 
Let 
\begin{eqnarray}
\!\!\!\!\!\!\!\!\Psi(\bar {\mathbf x},\rho) \!\!\!\!\!\!&\triangleq&\!\!\!\!\!\! -\! \sum_{i=1}^{K}\!\operatorname{ln}\biggl(\!-\biggl(\!1 \!-\!\Phi\biggl(\biggl[\frac{{\mathbf t}_{2i-1}^T\bar {\mathbf x}}{\sigma_{\tilde n}}, \;\frac{{\mathbf t}_{2i}^T\bar {\mathbf x}}{\sigma_{\tilde n}}\biggr]^T\!\!\!; \bar r\biggr)\!\!-\!\rho\biggr)\!\biggr)  \nonumber\\%\!\!-\! \operatorname{ln}(\rho)\nonumber\\
\!\!\!\!\!\!&&\!\!\!\!\! \!-\! \operatorname{ln}(-(\|\bar{\mathbf x}\| \!-\! \sqrt{P}))  %\nonumber\\ \!\!\!\!\!\!&&\!\!\!\!\! 
\!-\! \sum_{i=1}^{K}\!\operatorname{ln}(-(\|{\mathbf B}_{l}{\bar  {\mathbf x}}\| \!-\! \sqrt{\epsilon_l})) , 
\end{eqnarray} 
be the logarithmic barrier function. For $s >0$, define $\bar {\mathbf x}(s)$ and $\rho(s)$ as the solution of 
\begin{eqnarray}
\min_{\bar {\mathbf x},\rho} \; \rho + \Psi(\bar {\mathbf x},\rho)/s. \label{barrier}
\end{eqnarray}  
Problem (\ref {barrier}) is an unconstrained convex optimization problem and can be solved using the gradient descent algorithm \cite{boyd2004convex}. Problem (\ref{eq:worstser1}) is feasible if the constraints in (\ref{probconstrain})-(\ref{feasible2}) contain a non-zero feasible point. In particular, a feasible starting point of the barrier method can be computed as the solution of the following feasibility problem
\begin{eqnarray}
\max_{\!\bar  {\mathbf x},z} \; z \; \text{s.t.} \!\!\!\!\!&&\!\!\!\!\! z\!\leq\! {\mathbf t}_j^T \bar  {\mathbf x}, \; j\!=\!1,\!\dots,\!2K,\nonumber\\
\!\!\!\!\!&& \!\!\!\!\! \|\bar{\mathbf x}\| \!\leq\! \sqrt{P},\; \|{\mathbf B}_{l}{\bar  {\mathbf x}}\| \!\leq\! \sqrt{\epsilon_l} , \; l\!=\!1,\!\dots,\!L,\label{eq:feaprob}
\end{eqnarray}
which is a SOCP problem that can be solved efficiently. Suppose $(z^{\star},\bar{\mathbf x}^{\star}(1))$ is an optimal point of (\ref{eq:feaprob}).
The first observation is that if the assumption in (\ref{rhoineq}) is satisfied, then $z^{\star} \geq \sigma_{\tilde n_{j}}\alpha^{\star}(\bar r)$ is also satisfied. Reversely, if $z^{\star} < \sigma_{\tilde n_{j}}\alpha^{\star}(\bar r)$, then the assumption in (\ref{rhoineq}) is not true. In this case the problem is ill-posed as the SEP of the SUs exceed the values that are reasonable in practical applications. If $z^{\star} \geq \sigma_{\tilde n_{j}}\alpha^{\star}(\bar r)$, then, by Theorem $1$, problem (\ref{eq:worstser1}) is a convex problem and (\ref{eq:feaprob}) provides us a feasible starting point of the barrier method.
%If $z^{\star} < \sigma_{\tilde n_{j}}\alpha^{\star}(\bar r)$, then (\ref{eq:worstser1}) is infeasible.  

Algorithm \ref {algorithm1} summarizes the steps to compute the solution of (\ref{eq:worstser1}) using barrier method. For more details on the barrier method, the reader is referred to \cite[p.561-p.613]{boyd2004convex}.  We observe in the simulations that Algorithm \ref {algorithm1} provides a better performance compared to the conventional approach in \cite{Fu2009,Zhang2010}. 
\begin{algorithm}[htb]
	\caption{Efficient barrier method to solve (\ref{eq:worstser1})}
	\begin{algorithmic}
		\STATE \textbf{Given: } $s=1, \mu > 1$, tolerance $\delta_t>0$\\
		\STATE \textbf{Input: } $\{{\mathbf h}_i\}_{i=1}^{K}, \{b_i\}_{i=1}^{K}, P_0, \sigma, \bar r$ \\
		\STATE \textbf{Output: } The optimal solution $(\bar {\mathbf x}^{\star},\rho^{\star})$ of (\ref{eq:worstser1}) \\
		\STATE \text{Determine the optimal solution $(z^{\star},\bar{\mathbf x}^{\star}(1))$ of (\ref{eq:feaprob})}; \\
		\STATE \text{If $z^{\star} < \sigma_{\tilde n_{j}}\alpha^{\star}(\bar r)$, then no practical solution was found}; \\
		\STATE \text{Set} 
		\begin{eqnarray}
		\rho^{\star}(1) \!=\!\!\! \min_{1 \leq i \leq K} \!\{ 1 \!-\! \Phi(\tilde {\mathbf t}_{i}(\bar {\mathbf x}^{\star}(1));\bar r) \};  \nonumber
		\end{eqnarray} 	
		
		\REPEAT
		\STATE \text{Compute the optimal solution $(\bar {\mathbf x}^{\star}(s)$, ${\rho}^{\star}(s))$ of (\ref{barrier})}; \\
		\STATE \text{Set $s=\mu s$}; \\
		\UNTIL{$\|\bar  {\mathbf x}^{\star}(s)\! -\!\bar  {\mathbf x}^{\star}(s-1)\|^2\! + \!\|{\rho}^{\star}(s)\! -\!{\rho}^{\star}(s-1)\|^2\! <\! \delta_t^2$};\\ 
%		\IF {${\rho}^{\star}(s) > \rho_0$}
	%	\STATE \text{ (\ref{eq:worstser1}) is infeasible};
	%	\ELSE 
		\STATE \text{Output} $\bar  {\mathbf x}^{\star}(s), {\rho}^{\star}(s)$; 
	%	\ENDIF 
	\end{algorithmic}
	\label {algorithm1}
\end{algorithm}

\noindent{\it Remark : } The optimal beamforming vectors ${\mathbf w}^{\star}_i, i =1,...,K$ corresponding to the solutions of (\ref{eq:worstser}) can be computed using (\ref{eq:transmitted_signal}) as 
	\begin{eqnarray}
	{\mathbf w}^{\star}_i = \frac{{\mathbf x}^{\star} b_i^* }{K},
	\end{eqnarray}
where ${\mathbf x}^{\star}$ is the optimal solution in (\ref{eq:worstser}). 

\section{Computationally Efficient Approximate Approach} \label{section:ErrorProbability}
\subsection{Computationally Efficient Approximate WSUSEP Minimization Problem} 
In this subsection, we aim to provide a low complexity approximate approach to the WSUSEP-based CR downlink beamforming problem in (\ref{eq:worstser}) that achieves a tight approximation. Considering the addition law of probability, the left-hand side of (\ref{probconst1}) can be expressed as
	\begin{eqnarray}
	\operatorname{Pr}\bigl( {\psi_i} \in {\cal A}_{\theta }^{2\pi - \theta} \bigr)  
	\!\!\!\!&=&\!\!\!\! \operatorname{Pr}\bigl( {\psi_i} \in {\cal A}_{\theta }^{\pi + \theta} \bigr)
	%  \nonumber\\
 %\!\!\!\!&+&\!\!\!\!  
 + \operatorname{Pr}\bigl( {\psi_i} \in {\cal A}_{\pi - \theta}^{2\pi - \theta} \bigr)\nonumber\\
 \!\!\!\!&&\!\!\!\!- \operatorname{Pr}\bigl( {\psi_i} \in {\cal A}_{\pi - \theta}^{\pi + \theta}  \bigr),\label{eq:additionlaw}
	\end{eqnarray}
where 
	\begin{eqnarray}
	\operatorname{Pr}\bigl(\! {\psi_i} \in {\cal A}_{\theta }^{\pi + \theta} \bigr)\!\!\!\!&=&\!\!\!\!\operatorname{Pr}\bigl( {\psi_i} \in {\cal A}_{\theta }^{\frac{\pi}{2}}  \vee {\cal A}_{\frac{\pi}{2}}^{\pi + \theta}\bigr) \nonumber\\
	\!\!\!\!&=&\!\!\!\!\operatorname{Pr}\bigl(\tilde n_{2i-1} \geq {\mathbf t}_{2i-1}^T \bar  {\mathbf x} \bigr), \label{eq:probabilityconstraint3a}\\
	\operatorname{Pr}\bigl( \!{\psi_i} \in {\cal A}_{\pi - \theta}^{2\pi - \theta} \bigr) \!\!\!\!&=&\!\!\!\!
	\operatorname{Pr}\bigl( {\psi_i} \in {\cal A}_{\pi - \theta}^{\frac{3\pi}{2}}   \vee {\cal A}_{\frac{3\pi}{2}}^{2\pi - \theta}\bigr) \nonumber\\
	\!\!\!\!&=&\!\!\!\!\operatorname{Pr}\bigl( \tilde n_{2i} \geq {\mathbf t}_{2i}^T\bar  {\mathbf x} \bigr), \label{eq:probabilityconstraint3} %\label{eq:probcon1}
	\end{eqnarray}
are the probabilities that $\psi_i$ take values in the left and right half plane of Fig.~\ref{constellation}(b), respectively, i.e., between $\theta$ and $\pi+\theta$ and between $\pi-\theta$ and $2\pi-\theta$, respectively, $\vee$ is the logical ``OR" operator,  $\operatorname{Pr}\bigl({\cal A}_{\pi - \theta}^{\pi + \theta}  \bigr)$ is the probability of $\psi_i$ taking a value in the intersection of the left and right half planes given above. We remark that the beamformers are designed to steer the received signals into the corresponding corrected detection regions and hence generally $\operatorname{Pr}\bigl({\psi_i} \in {\cal A}_{\pi - \theta}^{\pi + \theta}  \bigr)$ takes small values. Therefore, by (\ref{eq:additionlaw}), (\ref{eq:probabilityconstraint3a}) and (\ref{eq:probabilityconstraint3}), we can approximate (\ref{probconst1}) as
	\begin{eqnarray}
	 \operatorname{Pr}\bigl({\psi_i} \in {\cal A}_{\theta }^{2\pi - \theta}\bigr)  
	 \!\!\!\!\!\!&\leq&\!\!\!\!\!\! \operatorname{Pr}\Bigl(\tilde n_{2i-1} \!\geq\! {\mathbf t}_{2i-1}^T\bar  {\mathbf x} \Bigr) \!+\! \operatorname{Pr}\Bigl( \tilde n_{2i} \!\geq\! {\mathbf t}_{2i}^T\bar  {\mathbf x} \Bigr)  \nonumber\\
	 \!\!\!\!\!\!&\leq&\!\!\!\!\!\! \rho.\label{eq:approx}
	\end{eqnarray}
Further restricting (\ref{eq:approx}) by the following two constraints
	\begin{eqnarray}
	\operatorname{Pr}\Bigl(\tilde n_{2i-1} \geq {\mathbf t}_{2i-1}^T\bar  {\mathbf x} \Bigr) \!  \leq \! \rho/2,\; \operatorname{Pr}\Bigl( \tilde n_{2i} \geq {\mathbf t}_{2i}^T\bar  {\mathbf x} \Bigr)\!  \leq \!\rho/2, \label{eq:approx2}
	\end{eqnarray}
the optimization problem in (\ref{eq:worstser}) can be approximately written as
\begin{subequations}	
	\label{eq:errorprob}
	\begin{eqnarray}
	\!\!\!\!\!\!\!\!\!\!\min_{\!\mathbf{x}, \rho} \!\!\!\!\!&&\!\!\!\!\!\! \rho \nonumber\\
	\!\!\!\!\!\!\!\!\!\!\text{s.t.} \!\!\!\!\!&& \!\!\!\!\!\! 
	\operatorname{Pr}\Bigl(\tilde n_{2i-1} \geq {\mathbf t}_{2i-1}^T\!\bar  {\mathbf x} \Bigr) \!  \leq \! \rho/2, , \; i\!=\!1,\!\dots,\!K, \label{eq:probabilityconstraint2a}\\
	\!\!\!\!\!&& \!\!\!\!\!\! \operatorname{Pr}\Bigl( \tilde n_{2i} \geq {\mathbf t}_{2i}^T\!\bar  {\mathbf x} \Bigr)\!  \leq \!\rho/2, \; i\!=\!1,\!\dots,\!K, \label{eq:probabilityconstraint2}\\
\!\!\!\!\!&& \!\!\!\!\!\! \|\mathbf{x}\|^2 \leq P, \|{\mathbf g}_{l}^T{\mathbf x}\|^2 \leq \epsilon_l ,  \; l\!=\!1,\!\dots,\!L,		
	\end{eqnarray}
\end{subequations}
which is the worst-case design on $\operatorname{Pr}\Bigl(\tilde n_{j} \geq {\mathbf t}_{j}^T\!\bar  {\mathbf x} \Bigr)$ for $j=1,\dots,2K$.
The approximate problem (\ref{eq:errorprob}) represents a restriction of problem (\ref{eq:worstser}) as 
in the sense that any optimal point of (\ref{eq:errorprob}) is feasible for (\ref{eq:worstser}), but the reverse statement is generally not true.
%(\ref{eq:approx2}) is satisfied then (\ref{eq:approx}) and therefore (\ref{probconst1}) must be satisfied. 
This means that (\ref{eq:errorprob}) is an inner approximation to (\ref{eq:worstser}).
%The constraints in (\ref{eq:probabilityconstraint2}) can be reformulated respectively as 
	%\begin{eqnarray}
	%	P_{2i-1}\!\! =&\operatorname{Pr}( \tilde n_{2i-1} \geq {\mathbf t}_{2i-1}^T\!\bar  {\mathbf x}) &\leq \rho/2, \label{eq:probabilityconstraint3a}\\
	%	P_{2i} =& \!\!\!\!\!\!\!\!\!\!\!\!\operatorname{Pr}( \tilde n_{2i} \geq {\mathbf t}_{2i}^T\!\bar  {\mathbf x}) &\leq \rho/2. \label{eq:probabilityconstraint3}
	%\end{eqnarray}
		%According to \cite{Chalise2007}, to ensure reliable communication link, we may assume that the error probability $p_j \leq 0.5$. 
		Based on (\ref{tildenoise}), we have \cite{Chalise2007}
		%The symbol error probability in (\ref{eq:probabilityconstraint3}) can be expressed as
		\begin{equation}
		\operatorname{Pr}\Bigl(\tilde n_{j} \geq {\mathbf t}_{j}^T\!\bar  {\mathbf x} \Bigr)=\int_{{\mathbf t}_{j}^T\!\bar  {\mathbf x}}^{\infty} \frac{\cos \theta}{\sqrt{\pi}\sigma}e^{\frac{\tilde n^2\cos^2 \theta}{\sigma^2}}d \tilde n. \label{eq:SEP}
		\end{equation}
		Using the Gaussian error function $\operatorname{erf}(\cdot)$, the SEP in (\ref{eq:SEP}) can be expressed as \cite{Chalise2007}
		\begin{equation}
		\operatorname{Pr}\Bigl(\tilde n_{j} \geq {\mathbf t}_{j}^T\!\bar  {\mathbf x} \Bigr)= \begin{cases}
		\frac{1}{2} + \frac{1}{2} \operatorname{erf}\Bigl(\frac{-{\mathbf t}_{j}^T\!\bar  {\mathbf x} \cos \theta}{\sigma} \Bigr), &  {\mathbf t}_{j}^T\!\bar  {\mathbf x} \leq 0, \\
		\frac{1}{2} - \frac{1}{2} \operatorname{erf}\Bigl(\frac{{\mathbf t}_{j}^T\!\bar  {\mathbf x} \cos \theta}{\sigma} \Bigr), &   {\mathbf t}_{j}^T\!\bar  {\mathbf x} \geq 0.
		\end{cases}
		\label{eq:probabilityconstraint4}
		\end{equation}
		Since $\operatorname{erf}(-x)=-\operatorname{erf}(x)$, we rewrite (\ref{eq:probabilityconstraint4}) as 
		\begin{equation}
		\frac{1}{2} - \frac{1}{2} \operatorname{erf}\Bigl(\frac{{\mathbf t}_{j}^T\!\bar  {\mathbf x} \cos \theta}{\sigma} \Bigr) \leq \rho/2,
		\label{eq:probabilityconstraint5}
		\end{equation}
		which is equivalent to
		\begin{eqnarray}
		\!\!\!\!\!\!\!\!\frac{\sigma\!\operatorname{erf}^{-1}(1\!-\!\rho)}{\cos \theta} \!\!\!\!\!&\leq&\!\!\!\!\!  \!{\mathbf t}_{j}^T\!\bar  {\mathbf x},
		\label{eq:probabilityconstraint6}
		\end{eqnarray}
		where $\operatorname{erf}^{-1}(\cdot)$ is the inverse error function. 
The approximate problem (\ref{eq:errorprob}) can be written as a function $\rho^{\star}(\cdot)$ for any given transmit power $P \geq 0$ such that
\begin{eqnarray}
\rho^{\star}(P):\min_{\!\bar  {\mathbf{x}}, \rho} \!\!\!\!\!&&\!\!\!\!\! \rho \nonumber\\
\!\!\!\!\!\! \text{s.t.} \!\!\!\!\!&& \!\!\!\!\! - {\mathbf t}_{j}^T\bar  {\mathbf x} + \frac{{\sigma\operatorname{erf}^{-1}\!(\!1\!-\!\rho)}}{\cos \theta} {\mathbf 1}_{2K} \leq 0, \; j\!=\!1,\!\dots,\!2K,\nonumber\\
\!\!\!\!\!&& \!\!\!\!\! \|\bar  {\mathbf x}\| \!\leq\! \sqrt{P},  \|{\mathbf B}_{l}{\bar  {\mathbf x}}\| \!\leq\! \sqrt{\epsilon_l} , \; l\!=\!1,\!\dots,\!L.	\label{eq:original3}
\end{eqnarray}
%where ${\mathbf T}$ is a $2K \times 2N$ matrix such that
% \begin{eqnarray}
% {\mathbf T}^T&\triangleq& [{\mathbf t}_1 \; {\mathbf t}_2 \; \dots \; {\mathbf t}_{2K} ].
% \label{eq:T}
% \end{eqnarray}
As the inverse error function $\operatorname{erf}^{-1}(v)$ is monotonously increasing in the interval $-1 < v < 1$, we can equivalently write (\ref{eq:original3}) as the following problem 
\begin{subequations}	
	\label{eq:robust}
\begin{eqnarray}
\!\!\!\!\!\!\!\!\max_{\!\bar  {\mathbf{x}}, \Upsilon} \!\!\!\!\!&&\!\!\!\!\!\! \Upsilon\sigma \nonumber\\
\!\!\!\!\!\!\!\! \text{s.t.} \!\!\!\!\!&&\!\!\!\!\!\! - {\mathbf t}_{j}^T\bar  {\mathbf x} \!+\! \frac{\Upsilon\sigma}{\cos \theta} {\mathbf 1}_{2K} \!\leq\! 0, \; j\!=\!1,\!\dots,\!2K, \label{eq:robustconst}\\
\!\!\!\!\!&&\!\!\!\!\!\! \|\bar  {\mathbf x}\| \!\leq\! \sqrt{P}, \|{\mathbf B}_{l}{\bar  {\mathbf x}}\| \!\leq\! \sqrt{\epsilon_l} , \; l\!=\!1,\!\dots,\!L,	
\end{eqnarray}
\end{subequations}
where $\Upsilon= \operatorname{erf}^{-1}(1 - \rho)$ is a scalar optimization variable and $\Upsilon^{\star}(\tilde P)$ is the optimal value of $\Upsilon$ in $(\ref{eq:optimalrelation})$ for a given transmit power $\tilde P$. Due to the monotonicity property of error function and inverse error function, there exists a one-to-one mapping from $\Upsilon$ to $\rho$, and vice versa.
Then we obtain the following relations:
\begin{eqnarray}
\Upsilon^{\star}(P)&=& \operatorname{erf}^{-1}(1-\rho^{\star}(P)),  \\
\rho^{\star}(P) &=& \frac{1}{2} - \frac{1}{2}\operatorname{erf}(\Upsilon^{\star}(P)),\\
\bar {\mathbf x}^{\star}_{\rho}(P)&=&\bar {\mathbf x}^{\star}_{\Upsilon}(P),
\label{eq:optimalrelation}
\end{eqnarray}
and $\bar  {\mathbf x}^{\star}_{\rho}(\tilde P)$ and $\bar {\mathbf x}^{\star}_{\Upsilon}(\tilde P)$ are optimal solutions of (\ref{eq:original3}) and (\ref{eq:robust}) for a given power $\tilde P$, respectively. Problem (\ref{eq:robust}) is SOCP and can be solved efficiently using convex optimization tools such as \texttt{CVX} \cite{grant2008cvx}.
The optimal beamforming vectors ${\mathbf w}^{\star}_i, i =1,...,K$ corresponding to the solutions of (\ref{eq:errorprob}) can be computed using (\ref{eq:transmitted_signal}) as 
	\begin{eqnarray}
	{\mathbf w}^{\star}_i = \frac{{\mathbf x}^{\star} b_i^* }{K},
	\end{eqnarray}
	where ${\mathbf x}^{\star}$ is the optimal solution in (\ref{eq:errorprob}).

\subsection{Computational Complexity}

		Here we provide a complexity comparison for the conventional and proposed approximate approaches, for the slow fading channel (respectively, fast fading channel). The conventional SINR balancing CR downlink beamforming problem	(\ref{eq:conventional1a}) is a sequential SOCP problem. The single SOCP in (\ref{eq:conventional1a}) can be solved with a worst-case complexity of $\mathcal{O}(N^3K^3(K+L)^{1.5})$ using efficient barrier methods \cite{Lobo1998}. The number of SOCP iterations is bounded above by $\mathcal{O}(\log I)$ where $I \delta_t$ is the range of the search space for $\gamma$ in (\ref{eq:conventional1a}) and $\delta_t$ is the error tolerance. As the optimization problem in (\ref{eq:conventional1a}) is data independent, it only needs to be applied once per frame (respectively, sub-frame) for the slow fading case (respectively, fast fading case). The complexity $C_{CA}$ per downlink frame (respectively, sub-frame) for the conventional CR approach is of the order
		\begin{eqnarray}
		C_{CA}\sim\mathcal{O}(N^3(K+L)^{1.5}K^3\log I). \label{CompCA}
		\end{eqnarray}
		The proposed approximate approach (\ref{eq:errorprob}) is a SOCP problem, which requires a worst-case complexity of $\mathcal{O}(N^3(K+L)^{1.5})$. As the proposed optimization problem in (\ref{eq:errorprob}) is data dependent, the number $N_{\rm SOCP}$ of SOCP compuations per frame for slow fading (respectively, the channel coherence sub-frame for fast fading) is equal to the number of data time-slots in the frame (respectively, sub-frame). Accordingly, the resulting complexity $C_{PAA}$ per downlink frame (respectively, sub-frame) for the proposed approximate approach is of the order
		\begin{eqnarray}
		C_{PAA}\sim\mathcal{O}(N^3(K+L)^{1.5}N_{\rm SOCP}). \label{CompPAA}
		\end{eqnarray}
		Therefore, if $\mathcal{O}(K^3\log I)$ and $\mathcal{O}(N_{\rm SOCP})$ are comparable, then it can be seen from (\ref{CompCA})-(\ref{CompPAA}) that the worst-case complexities of the proposed approximate and conventional CR downlink beamforming problems are also comparable. In the following simulations, using typical LTE Type 2 TDD frame scheme \cite{3GPP}, we show that the average execution time per downlink frames for the proposed approximate approach in (\ref{eq:errorprob}) is comparable to that of conventional CR beamforming for both slow and fast fading scenarios.
		
		It should be noted at this point that, compared to conventional beamforming, the proposed approaches provide significant complexity benefits at the receiver side. Indeed, as the received symbols lie at the constructive area of the constellation (see Fig.~\ref{constellation}), there is no need for equalizing the composite channel ${\mathbf h}_i^T{\mathbf w}_i^{c}$ to recover the data symbols at the $i$th SU,  where $\{{\mathbf w}_i^{c}\}_{i=1}^K$ is the optimal solution of (\ref{eq:conventional1a}). Accordingly, the benefit of the proposed approaches is that CSI is not required for detection at the SU. The benefit is in twofold aspect: 1.) There is no need for SBS to send common pilots for users to estimate the MISO channels. 2.) The SBS does not need to signal the beamformers for SUs compute the composite channels to decode the signals. Hence, the proposed approaches can save the training time and overhead for the signaling the beamformers for SUs, which leads to significant reductions in the operation time. %Nevertheless, in our schemes the second type of signaling (i.e., the signaling of the beamformers) would be prohibitive as it would have to be carried out at symbol level. 

 \subsection{Geometric interpretation}
  Problem (\ref {eq:robust}) can be interpreted as the problem of placing the centers of $K$ largest balls with centers $\bigl(\operatorname{Re}(b_i^*{\mathbf h}_i^T {\mathbf x}), \operatorname{Im}(b_i^*{\mathbf h}_i^T {\mathbf x})\bigr)$, ($i= 1,...,K$) and with maximum radii $ \Upsilon \sigma$ inside the decision region. This can then be interpreted as designing the SU beamformering vectors such that under a given power constraint the detection procedure applied to the received signal becomes most immune to noise as can be observed from Fig.~\ref{constellation}(c). The beamformer design maximizes the radius $\Upsilon\sigma$ of the noise uncertainty set (i.e., $\|n_i\| \leq \Upsilon\sigma$) within the correct detection region. As we can see from Fig.~\ref{constellation}(c), when the radius $\Upsilon\sigma$ of the noise uncertainty set is larger, the chance of the receive symbol falling outside the decision region reduce. As illustrated in Fig.~\ref{constellation}(c), %the quantity $\Upsilon \sigma$ defined in (\ref{eq:robust}) is the projection of ${\Upsilon \sigma }/{\cos \theta}$ from the imaginary axis. B
  by ensuring the correct detection region containing the noise uncertainty set, we have the following inequalities
 \begin{subequations}
 \begin{eqnarray}
\!\!\!\!\!\!\!\!\!\!\frac{|\operatorname{Im}(b_i^*{\mathbf h}_i^T {\mathbf x})|\! +\! {\Upsilon \sigma }/{\cos \theta}}{\operatorname{Re}(b_i^*{\mathbf h}_i^T {\mathbf x})} \leq\!\!\!\!&  \tan \theta, &\!\!\!\!\! \mbox{ for }  {\operatorname{Re}(b_i^*{\mathbf h}_i^T {\mathbf x})} \!>\! 0, \label{eq:real_contraints2}\\ 
\!\!\!\!\!\!\!\!\!\! \Upsilon =\!\!\!\!&  \!\!\!\!\!\!\!\!0,  &\!\!\!\!\!\mbox{ for } b_i^*{\mathbf h}_i^T {\mathbf x} \!=\! 0,  \label{eq:real_contraints3}
 \end{eqnarray}
 \end{subequations}
 which are equivalent to the constraint in (\ref{eq:robustconst}). 
 Therefore the proposed WSUSEP in (\ref{eq:worstser}) in its tight approximation in (\ref {eq:robust}) can also be interpreted as the following worst-users received symbol center placement problem:
 \begin{eqnarray}
 \max_{\!\mathbf{x}, \Upsilon} \!\!\!\!\!&&\!\!\!\!\! \Upsilon\sigma \nonumber\\
 \text{s.t.} \!\!\!\!\!&& \!\!\!\! \max_{\!\|n_i\| \leq \Upsilon\sigma\!}|\psi_i(\mathbf{x},n_i)| \leq \theta, \; i\!=\!1,\!\dots,\!K, \nonumber\\
 \!\!\!\!\!&& \!\!\!\!\! \|\mathbf{x}\|^2 \leq P, \; \|{\mathbf g}_{l}^T{  {\mathbf x}}\| \!\leq\! \sqrt{\epsilon_l} , \; l\!=\!1,\!\dots,\!L.
 \label{eq:original0}
 \end{eqnarray}

%\begin{figure}[ht]
%	\begin{minipage}[b]{1.0\linewidth}
%		\centering			 
%		\centerline{\epsfig{figure=n8m10M4va01run1000SERvsP.eps,width=9.5cm}}
%	\end{minipage}
%	\caption{SER performance versus power with $N=10$, $K=8$, and QPSK modulation.}
%	\label{fig1}
%\end{figure}

\begin{figure}[ht]
	\begin{minipage}[b]{1.0\linewidth}
		\centering			 
		\centerline{\epsfig{figure=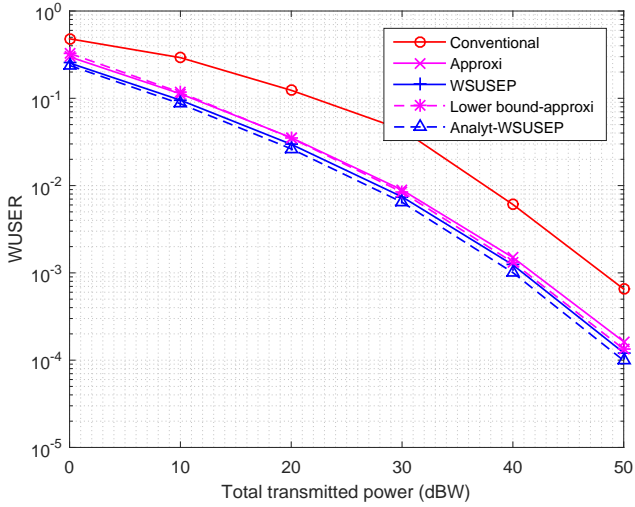,width=9.5cm}}
	\end{minipage}
	\caption{WUSER performance versus power with $N=10$, $L=2$, $K=8$, $\epsilon_l=-2 {\rm dBW}$, and QPSK modulation.}
	\label{fig2}
\end{figure}

%\begin{figure}[ht]
%	\begin{minipage}[b]{1.0\linewidth}
%		\centering			 
%		\centerline{\epsfig{figure=n8m10M4va01run1000BERvsP.eps,width=9.5cm}}
%	\end{minipage}
%	\caption{SER performance versus power with $N=10$, $K=8$, and QPSK modulation.}
%	\label{fig3}
%\end{figure}

% \begin{figure}[ht]
% 	\begin{minipage}[b]{1.0\linewidth}
% 		\centering			 
% 		\centerline{\epsfig{figure=n8m10M4va01run1000worstBERvsP.eps,width=9.5cm}}
% 	\end{minipage}
% 	\caption{WUBER performance versus power with $N=10$, $L=2$, $K=8$, $\epsilon_l=-2 {\rm dBW}$, and QPSK modulation.}
% 	\label{fig4}
% \end{figure}

 \begin{figure}[ht]
 	\begin{minipage}[b]{1.0\linewidth}
 		\centering			 
 		\centerline{\epsfig{figure=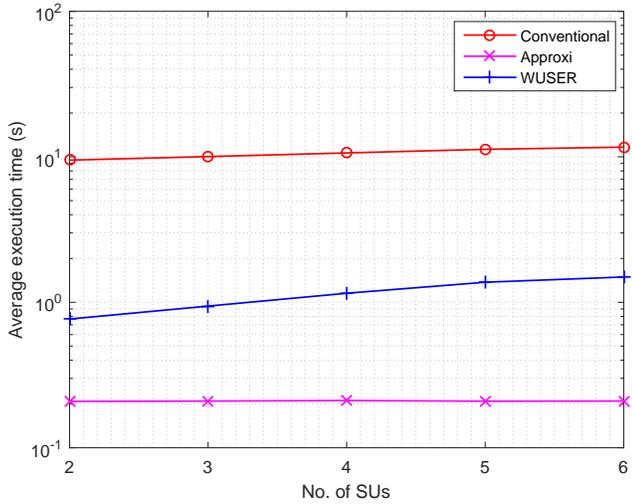,width=9.5cm}}
 	\end{minipage}
 	\caption{Average execution time per optimization versus number of SUs with $N=10$, $L=2$, $P=50 {\rm dBW}$, $\epsilon_l=-2 {\rm dBW}$, and QPSK modulation.}
 	\label{fig5}
 \end{figure}
 
  \begin{figure}[ht]
  	\begin{minipage}[b]{1.0\linewidth}
  		\centering			 
  			\vspace{-0.4cm}
  		\centerline{\epsfig{figure=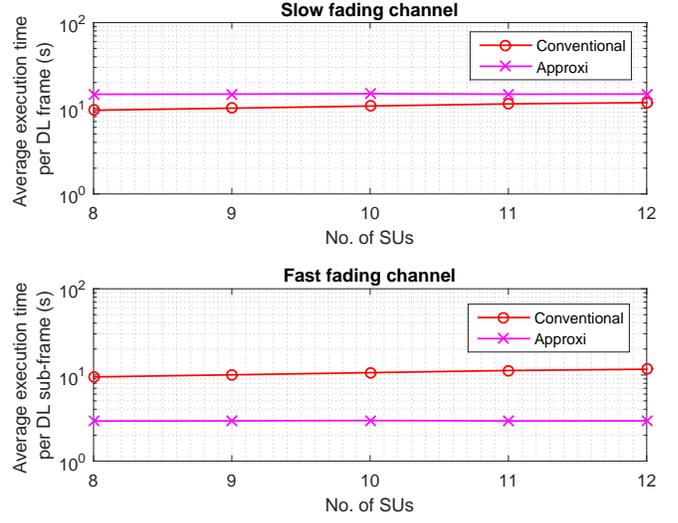,width=9.5cm}}
  	\end{minipage}
  	\caption{Average execution time versus number of SUs for slow/fast fading channels with $N=10$, $L=2$, $P=50 {\rm dBW}$, $\epsilon_l=-2 {\rm dBW}$, and QPSK modulation.}
  	\label{fig5a}
  \end{figure}
 
 \begin{figure}[ht]
 	\begin{minipage}[b]{1.0\linewidth}
 		\centering			 
 		\centerline{\epsfig{figure=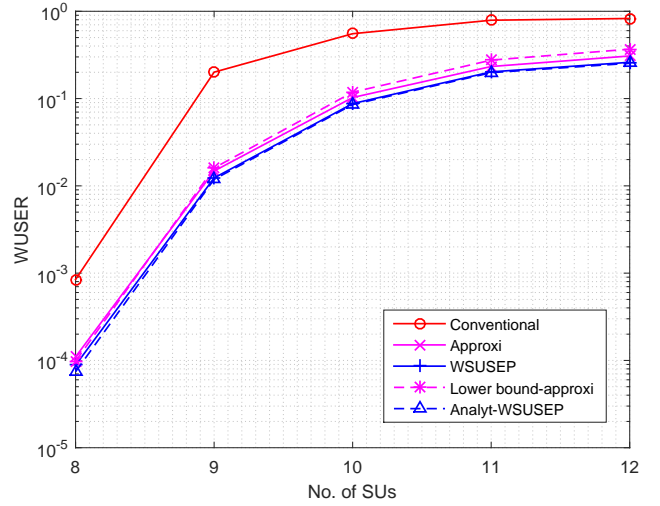,width=9.5cm}}
 	\end{minipage}
 	\caption{WUSER versus number of SUs with $N=10$, $L=2$, $P=50 {\rm dBW}$, $\epsilon_l=-2 {\rm dBW}$, and QPSK modulation.}
 	\label{fig6}
 \end{figure}

% \begin{figure}[ht]
% 	\begin{minipage}[b]{1.0\linewidth}
% 		\centering			 
% 		\centerline{\epsfig{figure=n4-8m10M4va01run1000worstBERvsN.eps,width=9.5cm}}
% 	\end{minipage}
% 	\caption{Worst user BER versus number of users with $N=10$, $P=5$dBW, and QPSK modulation.}
% 	\label{fig7}
% \end{figure}

\begin{figure}[ht]
	\begin{minipage}[b]{1.0\linewidth}
		\centering			 
		\centerline{\epsfig{figure=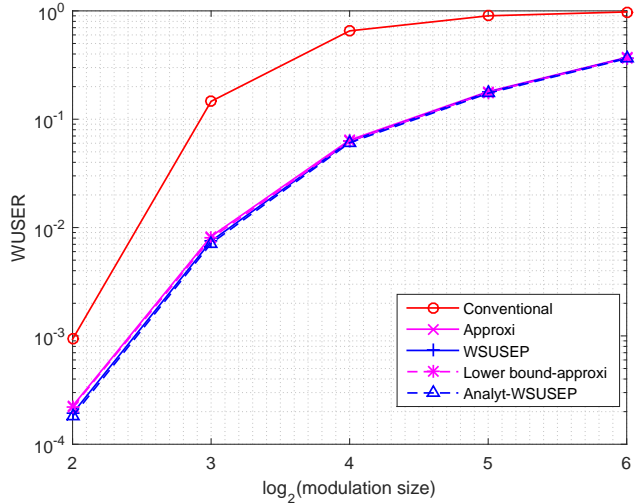,width=9.5cm}}
	\end{minipage}
	\caption{WUSER performance versus log of modulation size with $N=10$, $L=2$, $K=8$,$\epsilon_l=-2 {\rm dBW}$, and $P=50 {\rm dBW}$.}
	\label{fig8}
\end{figure}
\begin{figure}[ht]
	\begin{minipage}[b]{1.0\linewidth}
		\centering			 
		\centerline{\epsfig{figure=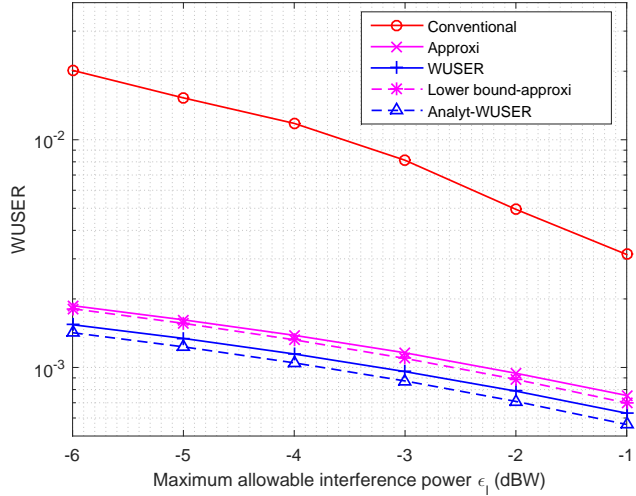,width=9.5cm}}
	\end{minipage}
	\caption{WUSER performance versus maximum allowable interference power $\epsilon_l$ with $N=10$, $L=2$, $K=8$, $P=40 {\rm dBW}$, and QPSK modulation.}
	\label{fig8a}
\end{figure}

%\begin{figure}[ht]
%	\begin{minipage}[b]{1.0\linewidth}
%		\centering			 
%		\centerline{\epsfig{figure=n8m10M4-64va01run1000worstBERvsMS.eps,width=9.5cm}}
%	\end{minipage}
%	\caption{Worst user BER performance versus log of modulation size with $N=10$, $K=8$, and $P=40$dBW.}
%	\label{fig9}
%\end{figure}

\begin{figure}[ht]
	\begin{minipage}[b]{1.0\linewidth}
		\centering
		\centerline{\epsfig{figure=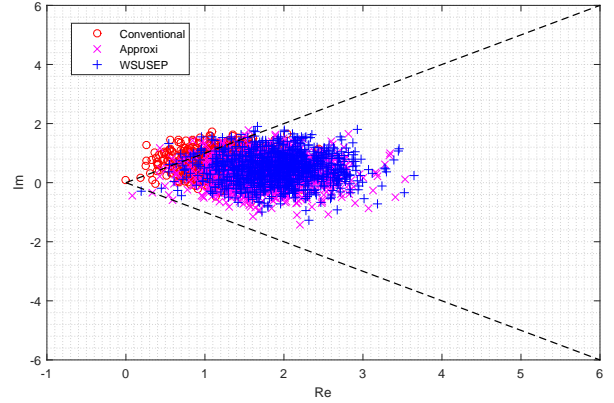,width=9.5cm}}
	\end{minipage}
	\caption{Distribution of received signals on complex plane where $N=10$, $L=2$, $K=8$, $\epsilon_l=-2 {\rm dBW}$, $P=5 {\rm dBW}$, and QPSK modulation.}
	\label{fig10}
\end{figure}

\begin{figure}[ht]
	\begin{minipage}[b]{1.0\linewidth}
		\centering
		\centerline{\epsfig{figure=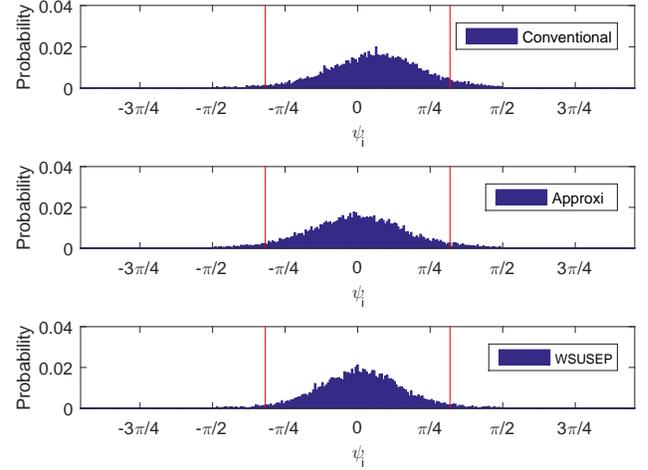,width=9.5cm}}
	\end{minipage}
	\caption{Histogram of the angles $\psi_i$ with $N=10$, $L=2$, $K=8$, $\epsilon_l=-2 {\rm dBW}$, $P=5 {\rm dBW}$, and QPSK modulation.}
	\label{fig11}
\end{figure}

\begin{figure}[ht]
	\begin{minipage}[b]{1.0\linewidth}
		\centering
		\centerline{\epsfig{figure=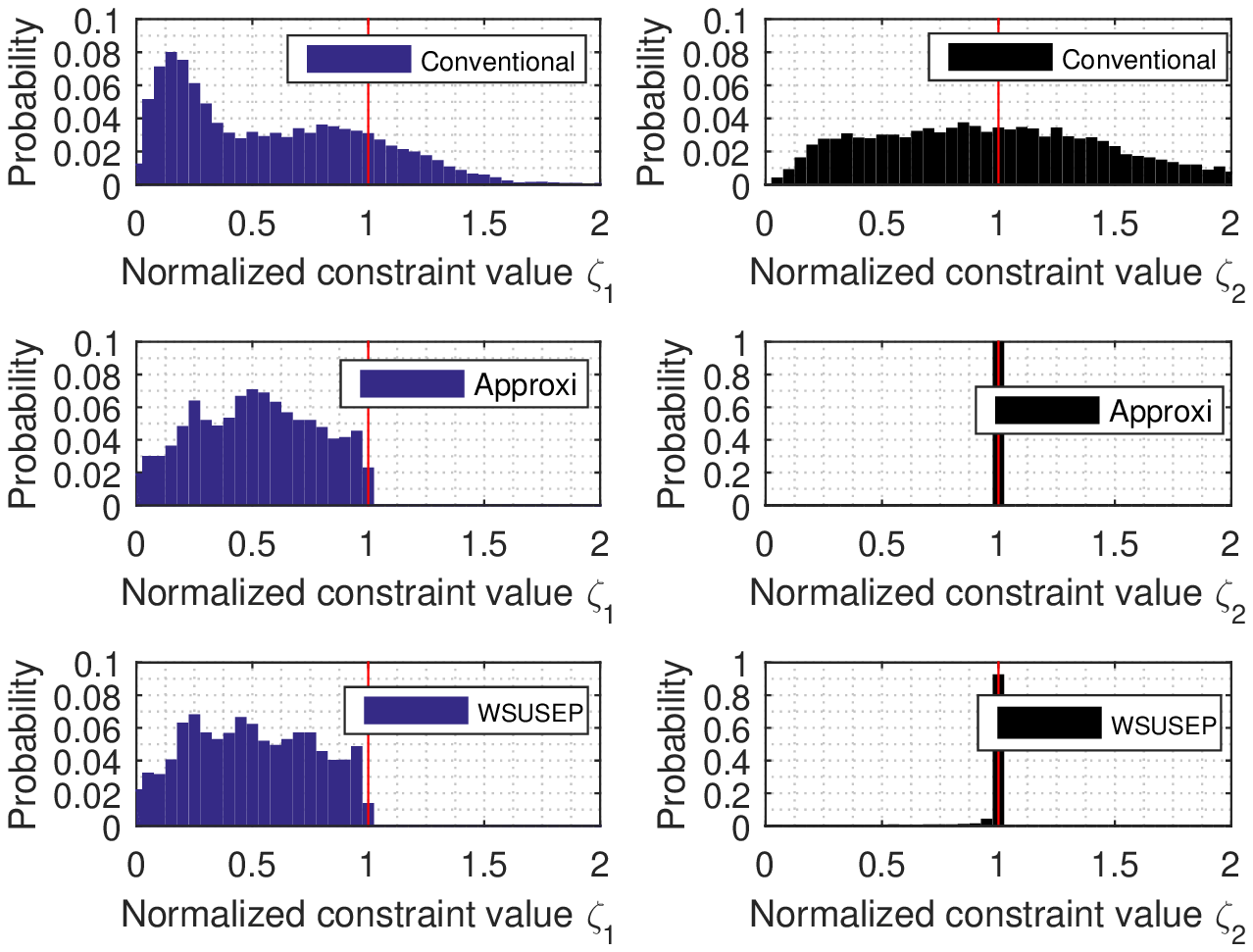,width=9.5cm}}
	\end{minipage}
	\caption{Histogram of normalized constraint values $\zeta_l$ with $N=10$, $L=2$, $K=8$, $\epsilon_l=-2 {\rm dBW}$, $P=5 {\rm dBW}$, and QPSK modulation.}
	\label{fig12}
\end{figure}

\section{Simulations}
In this section, we present simulation results for a constructive interference-based downlink beamforming for CR network with $N=10$ antennas and $L=2$ PU. Note that the benefits
	shown extend to different numbers of antennas. The system with $M$-PSK modulation is considered, i.e., $\theta=\pi/M$. A noise variance value of $\sigma^2=0.1$ is considered, while it is intuitive that the benefits of the proposed approach extend to other values. In line with \cite{huang2010rank}, we assume that the SUs and PUs connected to the SBS are located at directions  
\begin{eqnarray}
\!\!\!\!\!\!\!\!\omega_{1,...,10}\!\!\!\!\!\!&=&\!\!\!\!\!\!\begin{bmatrix}3^{\circ}\!\!, 35^{\circ}\!\!, 10^{\circ}\!\!, 39^{\circ}\!\!, 17^{\circ}\!\!, 74^{\circ}\!\!, 24^{\circ}\!\!, 86^{\circ}\!\!, 30^{\circ}\!\!, 80^{\circ}\end{bmatrix}^T \!+\! {\mathbf r}_1, \\
\!\!\!\!\!\!\!\!\tilde \omega_{1,2}\!\!\!\!\!\!&=&\!\!\!\!\!\![50^{\circ}\!, 57^{\circ}]^T \!+\! {\mathbf r}_2,
\nonumber %\label{eq:direction}
\end{eqnarray} 
 where ${\mathbf r}_1 \in \mathbb{C}^{10}$ and ${\mathbf r}_2 \in \mathbb{C}^2$ are drawn from a uniform distribution in the interval $[-1^{\circ}, 1^{\circ}]$. Then the downlink channel from the SBS to $i$th SU and  $l$th PU are modeled as \cite{huang2010rank} 
\begin{eqnarray}
{\mathbf h}_i \! &=\!&  \begin{bmatrix}1, e^{j\pi\sin\omega_i}, \dots, e^{j\pi(N-1)\sin\omega_i}\end{bmatrix}^{T}, \label{eq:channels}\\
{\mathbf g}_l\! &=\!&  \begin{bmatrix}1, e^{j\pi\sin\tilde \omega_l}, \dots, e^{j\pi(N-1)\sin\tilde \omega_l}\end{bmatrix}^{T}.
\end{eqnarray}
%We set $\epsilon_l=-2 {\rm dBW}$ for $l=1,2$. 
According to (\ref{psiangle}), we use the angle $\psi_i$ between the received signal $y_i$ and the transmitted symbol $b_i$ 
as an measure of the correct detection, which evaluates the performance of our proposed methods and the conventional method of (\ref{eq:conventional1a}). The receive signal can be correctly detected if $\psi_i$ is within the interval $[-\theta, \theta]$.
We introduce the normalized constraint value of interference power on an instantaneous basis \cite {Wajid2013}
\begin{align}
\zeta_l=\frac{\sum_{j=1}^K\sum_{i=1}^K{b}_j^*{b}_i{{\mathbf w}_j^{\star}}^H{\mathbf g}_l^*{\mathbf g}_l^T{\mathbf w}_i^{\star}}{\epsilon_l}, \label{normalizedconstraint}
\end{align}
as an abstract measure of the constraint satisfaction to compare the performance of different methods. The corresponding instantaneous interference power constraint at PU is satisfied if and only if $\zeta_l \leq 1$. %The error tolerance $\Delta_t=10^{-5}$ are used in Algorithm \ref{algorithm1}. For interior point methods, we set $\mu=60$. 
All results are average over $10^6$ Monte Carlo runs.

%As total power is the common metric between constructive interference-based and conventional optimizations, we focus on the optimization that can be compared by WUSER results and omit the power minimization approaches. 
In the following simulations, we compare three different techniques:
\begin{enumerate}
	\item  `Conventional' refers to the max-min fair problem (\ref{eq:conventional1a}) in \cite {Fu2009, Zhang2010};
	\item `WSUSEP' refers to the WSUSEP-based approach given in (\ref{eq:worstser1}); %use the interior point approach in Algorithm \ref{algorithm1} to solve (\ref{eq:wortser2a});
	\item `Approxi' stands for the computationally efficient approximation of the WSUSEP-based optimization given in  (\ref{eq:original3});
\end{enumerate}
and `Analyt-WSUSEP' and `Lower bound-approxi' stand for ${\rm mean}(\{\rho_i^{\star}\})$ and ${\rm mean}(\{\tilde \rho_i^{\star}\})$, respectively, where $\rho_i^{\star}$ and $\tilde \rho_i^{\star}$ are the optimal values of (\ref{eq:worstser1}) and (\ref{eq:original3}) for $i$th Monte Carlo run, respectively, and ${\rm mean}(\cdot)$ is the average function.  %Moreover, the value of `Analyt-sum SER' is the analytic SER. 

%Fig.~\ref{fig2}-~\ref{fig4} compare the performance versus the total power $P$ for the different techniques. 
In Fig.~\ref{fig2}, we fix the number of SUs and compare the performance of our proposed approaches and the conventional approach of (\ref{eq:conventional1a}) versus the total transmitted power $P$ for $K=8$, $\epsilon_l=-2 {\rm dBW}$, and QPSK modulation. It can be seen from Fig.~\ref{fig2} that the proposed approaches given in (\ref{eq:worstser1}) and (\ref{eq:original3}) outperform the conventional method of (\ref{eq:conventional1a}) in terms of the experimental WUSER performances. Notably, it can be observed in Fig.~\ref{fig2} that our analytic WUSER performance and lower bound of the computationally efficient approximate approach calculations match the experimental WUSER results of both (\ref{eq:worstser1}) and (\ref{eq:original3}), respectively. Furthermore, the computationally efficient approximate approach calculations match closely to the WSUSEP approach.
%In Fig.~\ref{fig4}, we investigate if the performance further improves for different techniques if channel code is applied. Here, we use the Gray code. Fig.~\ref{fig4} displays the worst user bit error rate (WUBER) versus the total transmitted power with $K=8$, $\epsilon_l=-2 {\rm dBW}$, when we set the modulation to be QPSK. Similarly, it can be noticed from the figure that the proposed approaches outperforms the conventional approach. Furthermore, we observe that the WUBER performance of all approaches in Fig.~\ref{fig4} reduce the error rate compared to the WUSER performance of our approaches in Fig.~\ref{fig2}. 
Note that in the simulations, we assume that in conventional approach the $i$th user receives the composite channel ${\mathbf h}_i^T{\mathbf w}_i^{c}$ from the SBS without any errors, which requires to decode the symbol $b_i$. 
However, the estimation of the composite channel ${\mathbf h}_i^T{\mathbf w}_i^{c}$ and feedback procedure to SU may result in estimation errors or reconstruction losses introduced by CSI quantization, which is required due to resource limitations of the feedback channels. The erroneous composite channel may further deteriorate the performances in Fig.~\ref{fig2} for the conventional approach. %-~\ref{fig4}

Fig.~\ref{fig5} compares the trend of the average execution time per optimization of our proposed methods and the conventional method for different number of SUs with $P=50 {\rm dBW}$, $\epsilon_l=-2 {\rm dBW}$, and QPSK modulation. As shown in Fig.~\ref{fig5}, when the number of SUs increase, the average execution time per optimization of the conventional method is much slower than the WSUSEP method and the computationally efficient approximate approach. Furthermore, we can see from the figure, the computationally efficient approximate approach is indifferent from the number of SUs. Note that we adopt the LTE Type 2 TDD frame scheme in \cite{3GPP}. 
Within a frame, 5 sub-frames, in which contains 14 symbols per each time-slot, are used for downlink (DL) transmission. Therefore, for the DL, it yields a block size of $B = 70$. A slow fading channel is assumed to be constant for the duration of one frame. In Fig.~\ref{fig5a}, we show the average execution time per DL frame for different number of SUs in slow fading channel scenario based on Fig.~\ref{fig5}. Although the end complexity is higher for our proposed method compared to the conventional method, the constructive interference based beamforming problem is still worthwhile as it improves the WUSER of SUs and secure the QoS of PUs on an instantaneous basis.
For fast fading channel, we assumed the channel to be constant for the duration of one sub-frame \cite{3GPP}. Fig.~\ref{fig5a} also depicts the average execution time per DL sub-frame for different number of SUs in fast fading channel scenario based on Fig.~\ref{fig5}. For this case, it can seen that our proposed approach offers a significant reduction in execution time down to around $30\%$. 

In Fig.~\ref{fig6}-~Fig.~\ref{fig8a}, we fix the transmitted power and vary the number of SUs, the size of PSK modulations, and the maximum allowable interference powers, respectively. Fig.~\ref{fig6} displays (both the experimental and analytic) WUSER performance for various number of SUs when we set the modulation to be QPSK with $\epsilon_l=-2 {\rm dBW}$ and $P=50 {\rm dBW}$. We observe that the experimental SER performance of our proposed approaches are better than the conventional approach. The computationally efficient approximate approach is a good approximation method for the WSUSEP approach. In Fig.~\ref{fig8}, we compare the performance versus different size of PSK modulations for the different techniques with  $K=8$,  $\epsilon_l=-2 {\rm dBW}$ and $P=50 {\rm dBW}$. As can be seen from the figure, the proposed methods outperform the conventional method especially when the size of modulations is small. Fig.~\ref{fig8a} depicts the WUSER performance versus different maximum allowable interference powers $\epsilon_l$ with $K=8$, $P=40 {\rm dBW}$, and QPSK modulation. We see from Fig.~\ref{fig8a} that our proposed methods perform better than the conventional method.

In Fig.~\ref{fig10}-~Fig.~\ref{fig12}, we look at the distribution of the received signals and instantaneous interference power, respectively. 
Fig.~\ref{fig10} depicts the distribution of the received signals using different techniques on complex plane with $K=8$, $\epsilon_l=-2 {\rm dBW}$, and $P=5 {\rm dBW}$, and QPSK modulation.
For the purposes of illustration, we show the example for which $b_i=1$ for $i=1,...,N$. Then we can see that the right side of dotted line is the correct detection region. In particular, the received signal is valid if it lays on the right side behind the dotted line. We observe from Fig.~\ref{fig10} that the received signals of our proposed methods can better fall into the correct detective region compared to the conventional method.
Fig.~\ref{fig11} displays the histograms of the angles $\psi_i$ between the received signal $y_i$ and the transmitted symbol $b_i$ with $K=8$, $\epsilon_l=-2 {\rm dBW}$, $P=5 {\rm dBW}$, and QPSK modulation. The angles outside the interval $[-\pi/4, \pi/4]$ are counted as errors. The first observation is that all approaches have normal-like distributions. As can be observed from Fig.~\ref{fig11}, there is relatively more outage for the conventional method compared to the proposed methods. Fig.~\ref{fig12} depicts the histograms of normalized constraint values $\zeta_l$ given in (\ref{normalizedconstraint}) with $K=8$, $\epsilon_l=-2 {\rm dBW}$, $P=5 {\rm dBW}$, and QPSK modulation. As can be observed from Fig.~\ref{fig12}, the conventional technique only satisfies about $50\%$ of the instantaneous interference power constraints for the second PU. This is due to the fact that the conventional method only considers the average interference power. However, our proposed approaches always satisfy the interference power constraints on an instantaneous basis. This consists of significant improvement over conventional CR beamformers which are prone to instantaneous PU outages.

\section{Conclusion} \label{sec:conclusion}
In this paper, we exploit the constructive interference in the underlay CR Z-channel by making use of CSI and transmit data information jointly. Our approach minimizes the WSUSEP of the SUs, subject to SBS transmit power constraints, while guaranteeing the PUs' QoS on an instantaneous basis. The proposed optimization can be formulated as a the bivariate probabilistic constrained programming problem. Under a condition on SEP, the problem can be expressed as a convex optimization problem and can be solved efficiently. We also propose a computationally efficient approximate approach to the WSUSEP approach to reduce the complexity. Simulation results have shown that our proposed methods have significantly improved performance as compared to the conventional CR downlink beamforming method. Future work can focus on extending our proposed constructive interference-based approaches by considering the robustness to CSI errors.

\appendix 

\subsection{Proof of Theorem $1$} \label{proofLMZ1}
%{\it Proof of Theorem $1$: }
%\begin{IEEEproof}
%Before proving the theorem, we formally define the bivariate normal CDF. 
To show the concavity, we need to use the first and second derivatives. 
It is well-known that taking the first derivative with respect to $u_1$, we have \cite{Prekopa1970}
\begin{eqnarray}
\frac{\partial\Phi({\mathbf u};r)}{\partial u_1}=\Phi({u_2 | u_1})\phi({u_1}),
\end{eqnarray}
where the conditional distribution function $\Phi({u_2 | u_1})$  is described by
\begin{eqnarray}
\Phi({u_2 | u_1})=\Phi\Bigl(\frac{u_2 - r u_1}{\sqrt{1 - r^2}}\Bigr).
\end{eqnarray}
Similarly, we have $\frac{\partial\Phi({\mathbf u};r)}{\partial u_2} \!= \Phi({u_1 | u_2})\phi({u_2})$.
Taking the second mixed derivative, we have
\begin{eqnarray}
\frac{\partial^2\Phi({\mathbf u};r)}{\partial u_1\partial u_2} \!&=& \!\phi\Bigl(\frac{u_2  \!- \! r u_1}{\sqrt{1  \!- \! r^2}}\Bigr)\frac{1}{\sqrt{1  \!- \! r^2}}\phi({u_1})\\
\!&=& \!\phi\Bigl(\frac{u_1  \!- \! r u_2}{\sqrt{1  \!- \! r^2}}\Bigr)\frac{1}{\sqrt{1  \!- \! r^2}}\phi({u_2}).
\end{eqnarray}
Taking the second derivative with respect to $u_1$, we have
\begin{eqnarray}
\frac{\partial^2\Phi({\mathbf u};r)}{\partial u_1^2} \!\!\!\!\!&=&\!\!\!\!\!\biggl(\!\phi\Bigl(\frac{u_2  \!- \! r u_1}{\sqrt{1  \!- \! r^2}}\Bigr)\frac{- \! r}{\sqrt{1  \!- \! r^2}} 
{- \! u_1}\Phi\Bigl(\frac{u_2  \!-  \!r u_1}{\sqrt{1  \!- \! r^2}}\Bigr)\!\biggr)\phi({u_1})\nonumber\\
\!\!\!\!\!&=&\!\!\!\!\!-r\frac{\partial^2\Phi({\mathbf u};r)}{\partial u_1\partial u_2}\!-\!u_1\frac{\partial\Phi({\mathbf u};r)}{\partial u_1}.
\end{eqnarray}
Similarly, we have $\frac{\partial^2\Phi({\mathbf u};r)}{\partial u_2^2} =-r\frac{\partial^2\Phi({\mathbf u};r)}{\partial u_1\partial u_2}\!-\!u_2\frac{\partial\Phi({\mathbf u};r)}{\partial u_2}$.
%\newtheorem*{Theorem2}{Theorem 2} 
%\begin{Theorem2}\cite{Prekopa1970}
%If $\eta_1,\eta_2$ have a joint normal distribution, with $-1 \leq r \leq 0$, where $\eta_1,\eta_2$ are standardized, then set of points satisfying the inequality
%\begin{eqnarray}
%\Phi(\mathbf u,r) \geq p,
%\end{eqnarray}
%is convex with respect to $u_1$ if $u_2$ is fixed and with respect to $u_2$ if $u_1$ is fixed where $p$ is a fixed probability satisfying the inequality 
%	\begin{eqnarray}
%	p \geq \Phi\biggl( \sqrt{\frac{\phi(1)}{2\Phi(1) + \phi(1)}} \biggr) \!\approx\! 0.6385.
%	\end{eqnarray}
%\end{Theorem2}
%\begin{IEEEproof}

By an abuse of notation, we redefine $\tilde n_j=\tilde n_j/\sigma_{\tilde n_j}$ and define $t_{j}(\bar {\mathbf x})\triangleq {\mathbf t}_{j}^T\bar {\mathbf x}/\sigma_{\tilde n_j}$. Then $\tilde n_j$ is a standardized random variable for all $j$.
%Moreover, It follows from \ref{prob1} that
%	\begin{eqnarray}
%	\Phi({\mathbf t}_{j}^T\bar {\mathbf x}) \geq \Phi({\mathbf t}_{2i\!-\!1}^T\bar {\mathbf x},{\mathbf t}_{2i}^T\bar {\mathbf x},r) \geq p,\; j=2i\!-\!1,2i,\; \forall i,
%	\end{eqnarray}
%	Thus, by \ref{p1}, we get
%	\begin{eqnarray}
%	{\mathbf t}_{j}^T\bar {\mathbf x} \!\geq\! \sqrt{\max_{v \geq 0} \frac{\phi(v)v}{\Phi(v)}} \!\approx\! 0.5427 ,\; \forall j=1,...,2K.
%	\end{eqnarray}
To show the standard bivariate normal CDF in (\ref{probconstrain}) is concave, it is enough to prove that the Hessian matrix of the CDF 
	\begin{eqnarray}
	\begin{bmatrix}{\frac{\partial {t}_{2i-1}}{\partial \bar {\mathbf x}}}^T \!\!\!&\!\!\!{\frac{\partial {t}_{2i}}{\partial \bar {\mathbf x}}}^T\end{bmatrix}{\mathbf M}_i \begin{bmatrix}{\frac{\partial {t}_{2i-1}}{\partial \bar {\mathbf x}}}^T \!\!\!&\!\!\!{\frac{\partial {t}_{2i}}{\partial \bar {\mathbf x}}}^T\end{bmatrix}^T,
	\label{eq:negdefine}
	\end{eqnarray}
is a negative-semidefinite matrix where 
	\begin{eqnarray}
	{\mathbf M}_i \triangleq\begin{bmatrix}\frac{\partial^2\Phi}{\partial t_{2i-1}^2}  &\frac{\partial^2\Phi}{\partial t_{2i-1}\partial t_{2i}}\\\frac{\partial^2\Phi}{\partial t_{2i-1}\partial t_{2i}} &\frac{\partial^2\Phi}{\partial t_{2i}^2}\end{bmatrix}.	
	\end{eqnarray}
The matrix in (\ref{eq:negdefine}) is negative-semidefinite if the eigenvalues $\lambda_i^{\pm}$ of ${\mathbf M}_i$ are negative, which are equal to
	\begin{eqnarray}
\lambda_i^{\pm} \!=\! \frac{\Bigl(\frac{\partial^2\Phi}{\partial t_{2i-1}^2} \!+\! \frac{\partial^2\Phi}{\partial t_{2i}^2}\Bigr) \pm \sqrt{\Delta_i}}{2},
	\end{eqnarray}
	where $\Delta_i\triangleq\Bigl(\frac{\partial^2\Phi}{\partial t_{2i-1}^2} \!-\! \frac{\partial^2\Phi}{\partial t_{2i}^2}\Bigr)^2 \!+\!4\Bigl(\frac{\partial^2\Phi}{\partial t_{2i-1}\partial t_{2i}}\Bigr)^2$. 
	First the eigenvalues are real values as $\Delta_i\geq 0$. Second, by (\ref{correlation}), we have $-1 \leq \bar r \leq 0$, for $M \geq 4$. 
	Then, by (\ref{p1}) and Lemma $1$B, we have
	\begin{eqnarray}
	\frac{\partial^2\Phi}{\partial t_j^2} \leq 0, \label{eq:secder}
	\end{eqnarray}
	which implies that $\Bigl(\frac{\partial^2\Phi}{\partial t_{2i-1}^2} \!+\! \frac{\partial^2\Phi}{\partial t_{2i}^2}\Bigr) \leq 0$. In order to show both eigenvalues are negative, we need to show that
	\begin{eqnarray}
	- \Bigl(\frac{\partial^2\Phi}{\partial t_{2i-1}^2} \!+\! \frac{\partial^2\Phi}{\partial t_{2i}^2}\Bigr) \geq \sqrt{\Delta_i},
\end{eqnarray}	
which is equivalent to
	\begin{eqnarray}
	\frac{\partial^2\Phi}{\partial t_{2i-1}^2} \frac{\partial^2\Phi}{\partial t_{2i}^2} \geq \Bigl(\frac{\partial^2\Phi}{\partial t_{2i-1}\partial t_{2i}}\Bigr)^2.
	\label{ineq1}
	\end{eqnarray}
If
	\begin{eqnarray}
	-\frac{\partial^2\Phi}{\partial t_{2i-1}^2}  \!\!\!\!\!&\geq&\!\!\!\!\! \frac{\partial^2\Phi}{\partial t_{2i-1}\partial t_{2i}}, \;%\label{ineq2}\\%
	-\frac{\partial^2\Phi}{\partial t_{2i}^2} \geq \frac{\partial^2\Phi}{\partial t_{2i-1}\partial t_{2i}},\label{ineq3}
	\end{eqnarray}
then (\ref{ineq1}) holds. The inequalities in (\ref{ineq3}) can be rewritten as %(\ref{ineq2}) and
\begin{eqnarray}
\frac{\Phi\Bigl(\frac{t_{2i} - \bar r t_{2i-1}}{\sqrt{1 - \bar r^2}}\Bigr) }{\phi\Bigl(\frac{t_{2i} - \bar r t_{2i-1}}{\sqrt{1 - \bar r^2}}\Bigr)}t_{2i-1} \!\!\!\!\!\!&\geq& \frac{1 - \bar r}{\sqrt{1 - \bar r^2}},\label{ineq4}\\
\frac{\Phi\Bigl(\frac{t_{2i-1} - \bar r t_{2i}}{\sqrt{1 - \bar r^2}}\Bigr) }{\phi\Bigl(\frac{t_{2i-1} - \bar r t_{2i}}{\sqrt{1 - \bar r^2}}\Bigr)}t_{2i} &\geq& \frac{1 - \bar r}{\sqrt{1 - \bar r^2}},
\label{ineq4a}
\end{eqnarray}
respectively. The inequalities in (\ref{ineq4}) and (\ref{ineq4a}) are satisfied for $t_{2i-1} \geq \alpha$, $t_{2i} \geq \alpha$, if we have 
\begin{eqnarray}
\frac{\Phi\Bigl(\alpha\frac{1- \bar r}{\sqrt{1 - \bar r^2}}\Bigr) }{\phi\Bigl(\alpha\frac{1 - \bar r }{\sqrt{1 - \bar r^2}}\Bigr)}\alpha \geq \frac{1 - \bar r}{\sqrt{1 - \bar r^2}}.
\label{ineq5}
\end{eqnarray}
To find the minimum $\alpha$, we can solve the optimization problem in (\ref{alphaopt}).	
%From (\ref{ineq5}), we can observe that $\alpha > 0$. Let $c \triangleq \frac{1 - \bar r }{\sqrt{1 - \bar r^2}}$. For $\alpha > 0$, the inequality in (\ref{ineq5}) can be written as
%\begin{eqnarray}
%	\frac{\Phi(\alpha c) }{\phi(\alpha c)} \alpha \geq c,
%	\label{ineq6}
%\end{eqnarray}
%where $c \geq 0$. Let $v= \alpha c$. Then (\ref{ineq6}) is equivalent to 
%\begin{eqnarray}
%\alpha^2 \geq \frac{\phi(v) v}{\Phi(v)}.
%\label{ineq7}
%\end{eqnarray}
%Note that $\frac{\phi(v) v}{\Phi(v)}$ is well defined for all $v \in \mathbb{R}$ and has first and second derivatives everywhere in $\mathbb{R}$, $\lim_{v \to \infty}\frac{\phi(v) %v}{\Phi(v)} = 0$ and $\lim_{v \to -\infty}\frac{\phi(v) v}{\Phi(v)} = -\infty$ by applying L'H\^opital's rule. 
%Moreover, we notice that when $v < 0$, then $\frac{\phi(v) v}{\Phi(v)}$ is also negative. Hence, to find the maximal point, we only need to consider when $v \geq 0$.
%By taking the first derivative, we can find the maximal point when 
%\begin{eqnarray}
%v^{\star}=-\frac{\phi(v^{\star})}{2\Phi(v^{\star})}+ \sqrt{\bigl(\frac{\phi(v^{\star})}{2\Phi(v^{\star})}\bigr)^2 +1} = 0.8399,
%\label{vopt}
%\end{eqnarray}
%with optimal value $\frac{\phi(v^{\star}) v^{\star}}{\Phi(v^{\star})}=0.2945$ and thus we can set the smallest $\alpha$ such that
%\begin{eqnarray}
%\alpha =\sqrt{\frac{\phi(v^{\star}) v^{\star}}{\Phi(v^{\star})}} \!\approx\! 0.5427.
%\label{alph1}
%\end{eqnarray}
Hence, for $t_{2i-1} \geq \alpha^{\star}(\bar r)$, $t_{2i} \geq \alpha^{\star}(\bar r)$, the inequalities in (\ref{ineq4}) and (\ref{ineq4a}) hold. This completes the proof of the theorem. $\hfill \blacksquare$

% Fig.\ref{fig13} depicts the surface of bivariate normal CDF with the correlation $r=0$, (i.e., when $M=4$). Fig.\ref{fig14} displays that when $u_1=u_2 \geq 0.5061$, the bivariate normal CDF $\Phi$ becomes concave. Note that when $u_1=u_2$, then $\Phi^2(u_1)=\phi({\mathbf u};r)$ and the inflection point of $\Phi^2(u_1)$ is at $u_1=\alpha^{\star}(r=0)\approx 0.5061$. This implies that the sufficient condition in (\ref{p1}) is also a tight condition. 
%\begin{figure}[ht]
%	\begin{minipage}[b]{1.0\linewidth}
%		\centering
%		\centerline{\epsfig{figure=bivariatecdf.eps,width=9.5cm}}
%	\end{minipage}
%	\caption{Bivariate normal CDF with $r=0$.}
%	\label{fig13}
%\end{figure}
%
%\begin{figure}[ht]
%	\begin{minipage}[b]{1.0\linewidth}%		
%		\centering
%		\centerline{\epsfig{figure=bivariatecdf1.eps,width=9.5cm}}
%	\end{minipage}
%	\caption{$\Phi$ vs $u_1$ with $u_1=u_2$. }
%	\label{fig14}
%\end{figure}

%\end{IEEEproof}

%\nocite{*}
\bibliographystyle{IEEEtran}	 
\bibliography{CIEfCRN}	

\end{document}